\documentclass{article}
\usepackage[dvips]{graphicx}
\usepackage{amssymb,amsmath,amsthm}

\newtheorem{lem}{Lemma}
\newtheorem{thm}[lem]{Theorem}

\begin{document}

\title{$6$-Cycle Double Covers of Cubic Graphs}

\author{Rodrigo S. C. Le\~ao\\
Valmir~C.~Barbosa\thanks{Corresponding author (valmir@cos.ufrj.br).}\\
\\
Universidade Federal do Rio de Janeiro\\
Programa de Engenharia de Sistemas e Computa\c c\~ao, COPPE\\
Caixa Postal 68511\\
21941-972 Rio de Janeiro - RJ, Brazil}

\date{}

\maketitle

\begin{abstract}
A cycle double cover (CDC) of an undirected graph is a collection of the graph's
cycles such that every edge of the graph belongs to exactly two cycles. We
describe a constructive method for generating all the cubic graphs that have a
$6$-CDC (a CDC in which every cycle has length $6$). As an application of the 
method, we prove that all such graphs have a Hamiltonian cycle. A sense of 
direction is an edge labeling on graphs that follows a globally consistent 
scheme and is known to considerably reduce the complexity of several distributed
problems. In \cite{mcsd}, a particular instance of sense of direction, called a 
chordal sense of direction (CSD), is studied and the class of $k$-regular 
graphs that admit a CSD with exactly $k$ labels (a minimal CSD) is analyzed. We 
now show that nearly all the cubic graphs in this class have a $6$-CDC, the only
exception being $K_4$. 

\bigskip
\noindent
\textbf{Keywords:}
Cycle double covers, Cubic graphs, Chordal sense of direction,
Circulant graphs, Hexagonal tilings, Fullerenes, Polyhexes.
\end{abstract}

\section{Introduction}
\label{intro}

In this paper we consider connected undirected graphs having no multiple edges
or self-loops. For terminology or notation not defined here we refer the reader
to \cite{bondy}. A \emph{cycle double cover} (CDC) of a graph $G$ is a
collection of cycles in $G$ such that every edge of $G$ belongs to exactly two
of the cycles. It can be easily seen that a necessary condition for a graph to
have a CDC is that the graph be $2$-edge-connected. It has been conjectured that 
this condition is also sufficient \cite{szekeres,seymour}, but the conjecture 
has remained unsettled and constitutes one of the classic unsolved problems in 
graph theory.

A \emph{$k$-cycle double cover} ($k$-CDC), for $k\geq 3$, is a CDC whose every
cycle has length $k$. Previous results on $k$-CDC's are the ones in
\cite{altshuler,thomassen2,thomassen1}, all motivated by the relationship 
between $k$-CDC's and embeddings on surfaces. In \cite{altshuler}, the 
$6$-regular graphs that have a $3$-CDC are studied and some results are shown to
carry over, by duality, to the class of girth-$6$ cubic graphs that have a $6$-CDC. 
This latter class is characterized in \cite{thomassen2}, and in \cite{thomassen1} 
deciding whether a graph has a $3$-CDC is proven NP-complete.

In this paper we introduce a constructive method for generating all the cubic
graphs that have a $6$-CDC, and prove in addition that all such graphs are 
Hamiltonian. For the particular case of girth-$6$ cubic graphs, these 
contributions provide both an alternative to the method of \cite{thomassen2} and 
an answer to the question raised in \cite{altshuler} regarding the graphs' 
Hamiltonicity. Graphs in this case can also be viewed as hexagonal 
tilings of the \emph{torus} (the orientable surface of genus $1$) or of the 
\emph{Klein bottle} (the non-orientable surface of cross-cap number $2$;
cf.\ \cite{thomassen2}) and have many applications in chemistry, where they are also 
called toroidal and Klein-bottle \emph{fullerenes} \cite{deza} or \emph{polyhexes} 
\cite{kirby}.

Our initial motivation, though, has been the relationship between $6$-CDC's and the 
\emph{chordal sense of direction} (CSD) \cite{flocchini8} of a graph. A sense of 
direction is an edge labeling on graphs that follows a globally consistent scheme 
and is known to considerably reduce the complexity of several distributed problems 
\cite{flocchini7}. 

In the particular case of a CSD, to be defined precisely in Section~\ref{mcsd},
we have in another study characterized the 
$k$-regular graphs that admit a CSD with exactly $k$ labels \cite{mcsd}, also 
called a \emph{minimal} CSD (MCSD). A further contribution of the present study is to 
demonstrate that, except for $K_4$, every cubic graph that has an MCSD also has a 
$6$-CDC. Since in \cite{mcsd} we also prove that the class of regular graphs that 
have an MCSD is equivalent to that of circulant graphs, this contribution also 
holds for cubic circulant graphs. We note that circulant graphs have great
practical relevance due to their connectivity properties (small diameter, high
symmetry, etc.), which render them excellent topologies for network
interconnection, VLSI, and distributed systems \cite{bermond}.

The following is how we organize the remainder of the paper. We start in Section~\ref{prelims}
with preliminary results on $6$-CDC's and their cycles. Then we 
move in Section~\ref{method} to the introduction of our method to generate all 
cubic graphs that have a $6$-CDC. In Section~\ref{applying} we describe the 
method's details by explaining how it is applied for each possible girth value. 
Our method never generates duplicates or misses a graph, as we explain in 
Section~\ref{complet} along with a demonstration that all cubic graphs that 
have a $6$-CDC are Hamiltonian. The relationship to MCSD's is discussed in 
Section~\ref{mcsd}, and then we close in Section~\ref{conclusions} with 
concluding remarks.   

\section{Preliminaries}
\label{prelims}

We henceforth assume that $G$ is a cubic graph on $n$ vertices and $m$ 
edges and that it has a $6$-CDC. Clearly, $m=3n/2$ and $n$ is necessarily even. 
We say that a cycle $C$ of such a $6$-CDC \emph{covers} a certain edge whenever that 
edge belongs to $C$. Let us initially establish some properties of the cycles of a 
$6$-CDC of $G$. 

\begin{lem}
\label{numcycles}
Every $6$-CDC of $G$ has $n/2$ cycles.
\end{lem}
\begin{proof} 
Let $t$ be the number of cycles in a $6$-CDC of $G$. As each edge belongs to two 
of the $t$ cycles, we have $6t=2m$. And because $m=3n/2$, it follows that 
$6t=3n$, thence $t=n/2$. 
\end{proof}

\begin{lem}
\label{nopath}
No two cycles of a $6$-CDC of $G$ share a path containing more than one edge. 
\end{lem}
\begin{proof}
Suppose, contrary to our aim, that $uvw$ is a path of length $2$ in $G$ 
belonging to two cycles, say $C$ and $C'$, of the $6$-CDC. Let $z$ be the other 
vertex adjacent to $v$. The two cycles of the $6$-CDC that cover the edge $vz$ 
must also cover either $uv$ or $vw$. However, both $uv$ and $vw$ are already 
covered by $C$ and $C'$, a contradiction. 
\end{proof}

\begin{lem}
\label{3cycles}
Every vertex of $G$ belongs to exactly three cycles of a $6$-CDC of $G$.
\end{lem}
\begin{proof}
By Lemma~\ref{nopath}, each of the cycles going through a vertex $v$ must cover 
a distinct pair of edges incident to $v$. The result follows from recognizing 
that there exist $\binom{3}{2}=3$ such pairs.
\end{proof}

Now, for $C$ and $C'$ any two cycles of a $6$-CDC of $G$, let $\mu(C,C')$ be the 
number of edges covered by both $C$ and $C'$, and $\sigma(C)$ the number of 
cycles of the $6$-CDC, excluding $C$, that have at least one edge in common with 
$C$. We can bound these numbers as follows.

\begin{lem}
\label{boundmu}
If $C$ and $C'$ are cycles in a $6$-CDC of $G$, then $0\leq \mu(C,C')\leq 3$.
\end{lem}
\begin{proof}
The lower bound is trivial and corresponds to $C$ and $C'$ being edge-disjoint. 
As for the upper bound, it follows directly from Lemma~\ref{nopath}, since 
$\mu(C,C')>3$ requires $C$ and $C'$ to share a path containing more than 
one edge.
\end{proof}

\begin{lem}
\label{boundsigma}
If $C$ is a cycle in a $6$-CDC of $G$, then $2\leq \sigma(C)\leq 6$.
\end{lem}
\begin{proof}
The lower bound follows from Lemma~\ref{nopath} and corresponds to $C$ having 
all three edges in each of its two possible sets of noncontiguous edges in 
common with a same cycle of the $6$-CDC. The upper bound is trivial and 
corresponds to the case in which $C$ shares each of its edges with a different 
cycle of the $6$-CDC.    
\end{proof}

\begin{figure}[t]
\centering
\begin{tabular}{c@{\hspace{1.5cm}}c}
\includegraphics[scale=0.65]{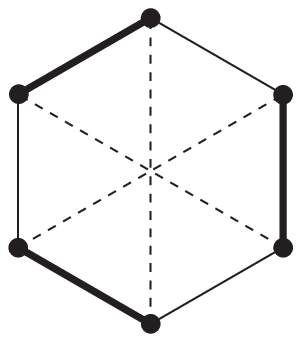}&
\includegraphics[scale=0.65]{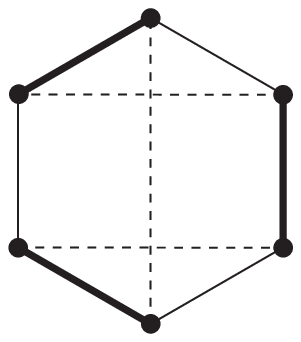}\\
{\small (a)}&{\small (b)}
\end{tabular}
\caption{The only cubic graphs that attain both the upper bound of
Lemma~\ref{boundmu} and the lower bound of Lemma~\ref{boundsigma}.
Each is shown with a three-cycle $6$-CDC highlighted (thin solid 
edges belong to one cycle, thick solid edges to another, dashed edges to both, 
and the external hexagon is the third cycle). The graph in part (a) is 
isomorphic to $M_6$ and the one in part (b) to $T_{6,2}$.}
\vspace{0.25in}
\label{matching}
\end{figure}

We can also characterize the graphs that effectively attain the upper bound of 
Lemma~\ref{boundmu}. As we see in Lemma~\ref{mu3} below, each such graph has a 
three-cycle $6$-CDC whose cycles all attain the lower bound of Lemma~\ref{boundsigma}
as well. We first review some definitions. A \emph{chord} is an 
edge interconnecting two noncontiguous vertices of a cycle. Let $C$ be a cycle 
of even length. A \emph{M\"obius ladder} on $n$ vertices, denoted by $M_n$, is 
the graph obtained by adding chords between all vertex pairs that are $n/2$ edges
apart on $C$ ($M_6$ is shown in Figure~\ref{matching}(a)). For $l$ a divisor of 
$n$, an \emph{$l$-layer torus} on $n$ vertices, denoted by $T_{n,l}$, is an 
$l \times n/l$ mesh in which maximally distant vertices on the same row or 
column are connected to each other ($T_{6,2}$ is shown in Figure~\ref{matching}(b)).

\begin{lem}
\label{mu3}
If $C$ and $C'$ are cycles in a $6$-CDC of $G$ such that $\mu(C,C')=3$, then $G$ is 
isomorphic to either $M_6$ or $T_{6,2}$.
\end{lem}
\begin{proof}
By Lemma~\ref{nopath}, it suffices to identify a perfect matching on one of $C$ 
or $C'$, say $C$, as the set of edges shared by the two cycles. The remaining 
edges of $C'$ can only be arranged so that $G$ is either isomorphic to $M_6$ 
or to $T_{6,2}$, as illustrated in Figure~\ref{matching}.
\end{proof} 

In the remainder of the paper, for any graph $H$ we use $V(H)$ to denote its 
vertex set and $E(H)$ its edge set. Furthermore, we call a \emph{cycle fragment} 
any subgraph of a $6$-CDC's cycle, and a \emph{cycle configuration} any 
collection of cycles or cycle fragments of a $6$-CDC. 

\section{A recursive method}
\label{method}

\begin{figure}[t]
\centering
\begin{tabular}{c@{\hspace{1.5cm}}c}
\includegraphics[scale=0.65]{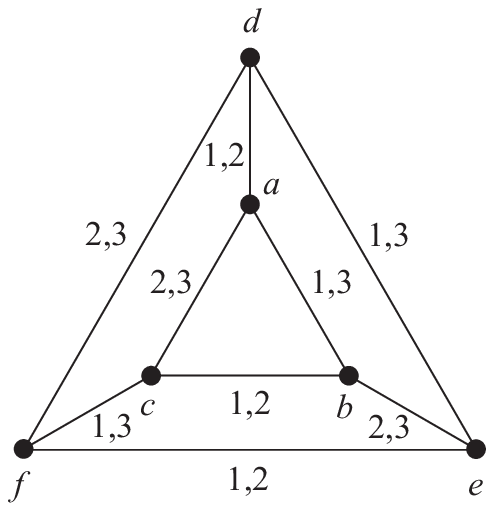}&
\includegraphics[scale=0.65]{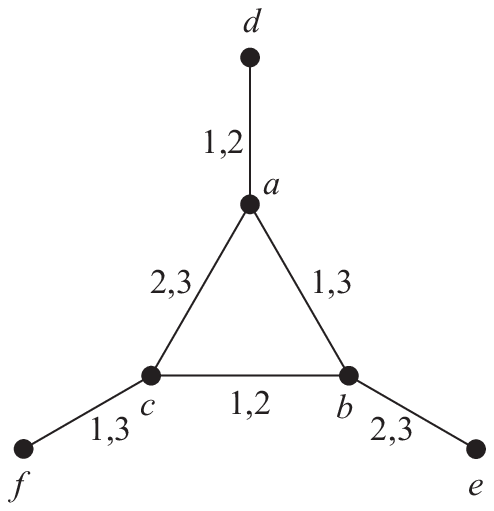}\\
{\small (a)}&{\small (b)}
\end{tabular}
\caption{Graph $G$ with its $6$-CDC (a) and a subgraph $H$ of $G$ with the
corresponding cycle fragments from the $6$-CDC of $G$ (b).}
\vspace{0.25in}
\label{b3&s3}
\end{figure}

Having established some restrictions on the cycles that form a $6$-CDC, we are
now in position to describe a constructive method for generating cubic graphs
that have a $6$-CDC. Let $H$ be a proper subgraph of $G$ whose vertices have degree
$1$ or $3$. When we consider the intersection of the cycles of a $6$-CDC of $G$
with $H$, we obtain a cycle configuration in $H$ such as the one illustrated in
Figure~\ref{b3&s3}, where an edge is labeled $i,j$ to indicate that it belongs
to the cycles $C_i$ and $C_j$ of the $6$-CDC.

We say that a vertex is \emph{deficient} in a certain graph if its degree is
less than $3$ in that graph. The \emph{deficiency} of graph $H$ is given by
$3\vert V(H)\vert-\sum_{v\in V(H)}{d_H(v)}$, 
where $d_H(v)$ is the degree of $v$ in $H$. Note that the deficient vertices in 
$H$ are exactly the ones that are also deficient in $G-E(H)$, where for 
$E\subset E(G)$ we use $G-E$ to denote the graph obtained from $G$ by removing 
the edges in $E$ and the vertices that become isolated after the edge removal. For 
$G$ and $H$ as in Figure~\ref{b3&s3}, $G-E(H)$ is the external triangle in part 
(a) of the figure.

\begin{figure}[t]
\centering
\begin{tabular}{c@{\hspace{0.8cm}}c}
\includegraphics[scale=0.65]{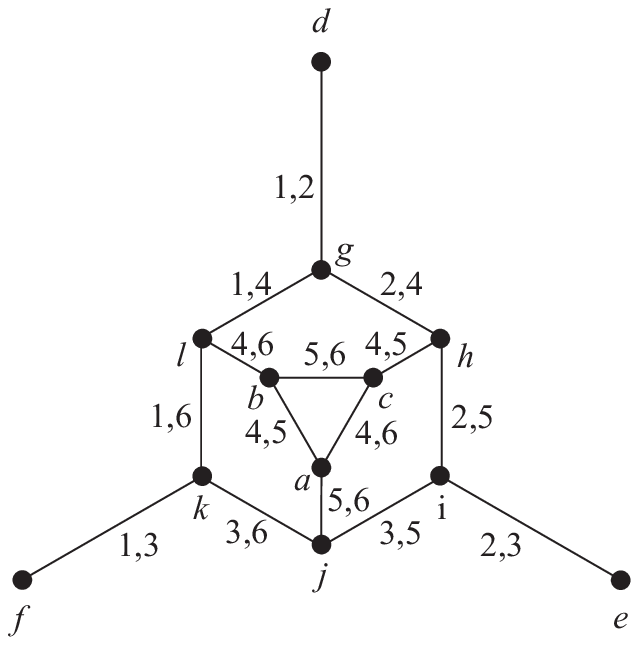}&
\includegraphics[scale=0.65]{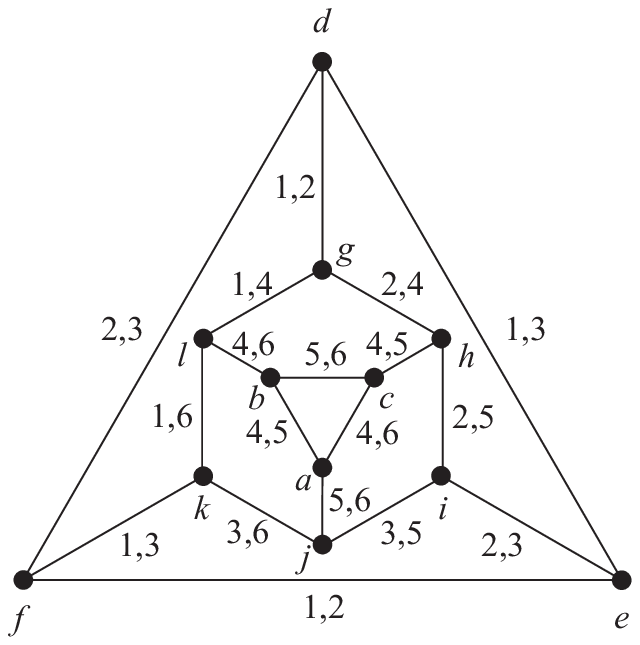}\\
{\small (a)}&{\small (b)}
\end{tabular}
\caption{Graph $H'$ (a) equivalent to the $H$ of Figure~\ref{b3&s3}(b) and 
the resulting $G'$ (b).}
\vspace{0.25in}
\label{I3&1iter}
\end{figure}

Now suppose that there exists another graph $H'$ that can replace $H$ in $G$ in
such a way that the resulting graph, call it $G'$, is cubic and has a $6$-CDC
with the same cycle configuration on the edges of $G'-E(H')$ as on the edges of
$G-E(H)$. In other words, suppose that $H'$ leads to a cubic
$G'=H'\cup [G-E(H)]$ such that the cycle configuration in $G-E(H)$ remains
unchanged from the $6$-CDC of $G$ to that of $G'$. If such is the case, then
$H'$ must have the same deficiency as $H$ (so the resulting $G'$ is cubic) and
also a cycle configuration that completes the one in $G-E(H)$ as needed to yield
the $6$-CDC of $G'$. For $G$ and $H$ as in Figure~\ref{b3&s3}, $H'$ and $G'$ are
as illustrated in Figures~\ref{I3&1iter}(a) and \ref{I3&1iter}(b), respectively.

For the latter condition to be satisfied, the following two properties must
hold. First, if $v_1,\ldots,v_y$ are the deficient vertices of $H$ (all of
degree $1$, by definition), then there has to exist a partition
$\{U_1,\ldots,U_y\}$ of the deficient vertices of $H'$ such that $v_x$ and $U_x$
have the same deficiency,\footnote{We extend the definition of a graph's
deficiency to that of a vertex or vertex set in the obvious way.}
$x=1,\ldots,y$. Clearly, $\vert U_x\vert\le 2$ necessarily, so $U_x$ has either
one degree-$1$ vertex or two degree-$2$ vertices. By Lemma~\ref{3cycles}, each
degree-$1$ vertex in $H$ or $H'$, or degree-$2$ vertex in $H'$, has exactly two
cycles of the $6$-CDC of $G$ going through it along edges of both $H$ and
$G-E(H)$, or of $H'$ and $G'-E(H')$, as the case may be. For $v_x$ and $U_x$,
$x=1,\ldots,y$, let $C_i$ and $C_j$ be the two cycles in $H$, $C_i'$ and $C_j'$
the two cycles in $H'$. The second property is that the fragment of $C_i$ in $H$
and the fragment of $C_i'$ in $H'$ both have the same length, and similarly for
$C_j$ and $C_j'$.

In the case of Figures~\ref{b3&s3} and \ref{I3&1iter}, no
degree-$2$ vertices exist in $H'$ and the above holds
with $C_i=C_i'=C_1$ and $C_j=C_j'=C_2$, for example. The fragments of 
$C_1$ and $C_2$ in $H$ are $dabcf$ and  $dacbe$, respectively, while in $H'$ they 
are $dglkf$ and $dghie$.

We say that cycle configurations such as the ones of $H$ and $H'$ are 
\emph{equivalent} to each other. When it is the case, in addition, that $H'$ has 
at least one subgraph that is isomorphic to $H$ and all such subgraphs have cycle 
configurations that are equivalent to that of $H$ (hence to that of $H'$ also), 
then we say that $H'$ is \emph{self-similar} with respect to $H$. This is 
certainly the case of the $H$ and $H'$ of Figures~\ref{b3&s3}(b) and 
\ref{I3&1iter}(a), since the triangle of Figure~\ref{I3&1iter}(a), when 
augmented by vertices $h$, $j$, and $l$ and the edges that lead to them from 
the triangle, is isomorphic to the graph in Figure~\ref{b3&s3}(b) with 
equivalent cycle configuration.      

Self-similarity is a property of positive-deficiency graphs and constitutes the 
core of our method. Before proceeding to a description of the method, 
we let $g(G)$ denote the girth of $G$.
In our present case of graphs that have a $6$-CDC, 
$3\leq g(G)\leq 6$ necessarily, and we use the value of $g(G)$ to divide our 
approach into cases, as presented in Section~\ref{applying}.

For a fixed value of $g$ in $\{3,\ldots,6\}$, let $S_g$ be the deficiency-$2g$ 
graph on $2g$ vertices that comprises a length-$g$ cycle and $g$ additional 
vertices, each of them connected to a distinct vertex of the cycle. For $g=3$, 
$S_g$ is the $H$ of Figure~\ref{b3&s3}(b). In general, it is easy to see that
every girth-$g$ cubic graph having a $6$-CDC has a subgraph isomorphic to $S_g$
with a cycle configuration that is consistent with the $6$-CDC,
even though the $g$ off-cycle vertices of this
subgraph are not always all distinct.\footnote{In fact, vertex distinctness holds
for all but one single case, specifically one of the cycle configurations of $S_4$,
as we discuss in Section~\ref{girth4}.} For a fixed cycle configuration of 
$S_g$, let also $I_g$ be a minimal girth-$g$ proper supergraph of $S_g$ which, 
along with a cycle configuration of its own, is self-similar with respect to 
$S_g$.\footnote{Notwithstanding the formal generality of this definition, what happens
is that, as we show in Sections~\ref{girth3} through \ref{girth6}, for every valid
cycle configuration of $S_g$ there exists only one $I_g$ instance.} In the $g=3$
example, $I_g$ is the $H'$ of Figure~\ref{I3&1iter}(a).

Now, a very important observation is that, for $g=6$, it may be impossible for $I_g$
to exist as defined. The reason is that the cycle configuration of $S_6$ does not
necessarily include a complete cycle of the $6$-CDC, while it may happen that every
girth-$6$ proper supergraph of $S_6$ whose cycle configuration is equivalent to that
of $S_6$, is also a supergraph of an isomorph of $S_6$
whose cycle configuration does include a complete $6$-CDC cycle. We then see that the
definition of self-similarity must be modified in the girth-$6$ case when the cycle
configuration of $S_6$ does not contain a complete $6$-CDC cycle. The modification
is that not all
subgraphs of $I_6$ that are isomorphic to $S_6$ are required to have cycle configurations
equivalent to that of $S_6$, but rather only those whose cycle configurations do not contain a
complete $6$-CDC cycle. That the girth-$6$ case should require such an exceptional treatment
is not really a surprise, since we are throughout dealing with graphs that have a
$6$-CDC, and thence it is only natural that $6$-cycles that are in the $6$-CDC be
distinguished from those that are not.

Given the notion of self-similarity, the definitions of $S_g$ and $I_g$ imply 
that $I_g$ can substitute indefinitely for any isomorph of $S_g$ that has a cycle
configuration equivalent to that of $I_g$, thus generating an infinite sequence of
deficiency-$2g$, girth-$g$ graphs whose first graph is $S_g$ itself. For $g<6$,
such an isomorph is any of the $S_g$-isomorphs that the current graph has as subgraphs;
for $g=6$, isomorphs whose cycle configuration includes a complete $6$-CDC cycle are
excluded if the cycle configuration of $S_g$ does not itself contain a complete $6$-CDC
cycle. Turning the resulting graphs into girth-$g$ cubic graphs that have a $6$-CDC
requires that we define yet another graph based on $S_g$.
This graph is denoted by $B_g$ and its definition, too, depends on what happens in the
$g=6$ case.

$B_g$ is in all cases defined to be a girth-$g$ supergraph of $S_g$ that
has a $6$-CDC.
If either $g<6$ or else $g=6$ but the cycle configuration of $S_g$ does not
include a complete $6$-CDC cycle, then $B_g$
is furthermore of one of two types:
\begin{enumerate}
\item[(i)] $B_g$ is not a supergraph of $I_g$.
\item[(ii)] $B_g$ is a supergraph of $I_g$ such that
substituting $S_g$ for $I_g$ causes the girth of $B_g$ to be reduced.
\end{enumerate}
The remaining case is that of $g=6$ when
the cycle configuration of $S_g$ does include a complete cycle of the $6$-CDC.
In this case, $B_g$ has the following property, in addition to being a girth-$g$
supergraph of $S_g$ that has a $6$-CDC:
\begin{enumerate}
\item[(iii)] $B_g$ is a supergraph of $I_g$ and all its $6$-cycles are cycles of the
$6$-CDC. In addition, it is such that substituting $S_g$
for $I_g$ causes the appearance of $6$-cycles that are not in the $6$-CDC.
\end{enumerate}
In any of cases (i)--(iii), the $6$-CDC of $B_g$ is assumed to coincide with the cycle 
configuration of $S_g$ or $I_g$, depending respectively on whether $B_g$ is a 
supergraph of $S_g$ only or of $I_g$ as well. It is also curious to note that,
if $B_g$ is of type (iii), then in $I_g$ it automatically holds that every $6$-cycle
is a cycle of the $6$-CDC; but this already follows from
the very definition of $I_g$, since in this case $S_g$ itself contains a complete $6$-CDC
cycle in its cycle configuration.

The reason for making a distinction between these three types is immaterial at this 
point and will only become clear in Section~\ref{girth6}, in which we handle the
girth-$6$ case, and in Section~\ref{complet} when we argue for the 
completeness of our method. For the girth-$3$ example we have been using as 
illustration, notice that the $G$ of Figure~\ref{b3&s3}(a) is a type-(i) instance 
of $B_3$. 

One crucial property emerging from the definitions of $S_g$, $I_g$, and $B_g$ is 
that, except for type-(i) instances of $B_g$, every girth-$g$ cubic graph that has 
a $6$-CDC also has a subgraph isomorphic to $I_g$ with a cycle 
configuration that renders it self-similar with respect to $S_g$.
So not only is the indefinite substitutability of $I_g$ for $S_g$ 
true, but it can be used to generate all girth-$g$ cubic graphs that have a 
$6$-CDC, as follows. For each possible cycle configuration of $S_g$, we identify 
$I_g$ and all pertinent $B_g$ instances. By starting at each such instance and 
substituting $I_g$ for $S_g$ indefinitely, ever larger girth-$g$ cubic graphs are 
generated having a $6$-CDC.      

\section{Applying the method}
\label{applying}

For each pertinent girth value $g$, in this section we start with $S_g$ and 
identify all its possible cycle configurations. For each of these cycle 
configurations, we then expand $S_g$ (along with its cycle configuration, by 
adding vertices and edges) without disrupting the $6$-CDC nature of its cycle 
configuration or altering the girth. We do this until $I_g$ and all instances of 
$B_g$ are obtained.

While expanding $S_g$ we first attempt to generate type-(i) instances of $B_g$,
that is, those that are not supergraphs of $I_g$. Then we proceed to generating
$I_g$ itself and from there we move to expanding $I_g$ towards obtaining the
instances of $B_g$ that are supergraphs of $I_g$, that is, type-(ii) or (iii)
instances. It is important to realize that, since $S_g$ and $I_g$ have cycle
configurations that are equivalent, carrying the expansion beyond $I_g$ need
not attempt the same expansion steps that generated type-(i) $B_g$ instances:
doing this would only lead to graphs that already belong to the sequence of graphs
generated by substituting $I_g$ for $S_g$ recursively from a type-(i) $B_g$ instance
onward. What must be attempted, rather, are expansion steps that failed
previously but may now succeed (like those that somehow disrupt the girth
or the $6$-CDC when attempted on $S_g$).

\subsection{The girth-$\mathbf{3}$ case}
\label{girth3}

We start with the graph of Figure~\ref{b3&s3}(b) as $S_3$ (that is, the core cycle
of $S_3$ is $abca$). It is easy to see that the cycle configuration given in 
Figure~\ref{b3&s3}(b) is the only one that does not violate the restrictions 
discussed in Section~\ref{prelims}. Furthermore, note that the vertices 
$d$, $e$, and $f$ must all remain distinct as we expand $S_3$, otherwise either 
the resulting graph would be isomorphic to $K_4$ (which is too small to have a 
$6$-CDC) or the resulting cycle configuration would be inconsistent with the 
requirements of a $6$-CDC.

Given the unique cycle configuration for $S_3$ in Figure~\ref{b3&s3}(b), we 
proceed with the expansion. This is done by completing the cycles $C_1$, $C_2$, 
and $C_3$. We have two ways of completing $C_1$: either using an existing vertex
($e$) or including a new one. While the former option leads unavoidably to $B_3$
and its cycle configuration shown in Figure~\ref{b3&s3}(a) when applied to 
all three cycles, the latter results, after a suitable renaming of vertices and 
cycles, and also unavoidably, in the $I_3$ of Figure~\ref{I3&1iter}(a) and its 
cycle configuration. As noted in Section~\ref{method}, $B_3$ is of type (i);
also, for the reasons given above in the introduction to Section~\ref{applying},
seeking type-(ii) instances of $B_3$ any further is in this case meaningless.

\begin{figure}[t]
\centering
\begin{tabular}{c@{\hspace{1.5cm}}c}
\includegraphics[scale=0.65]{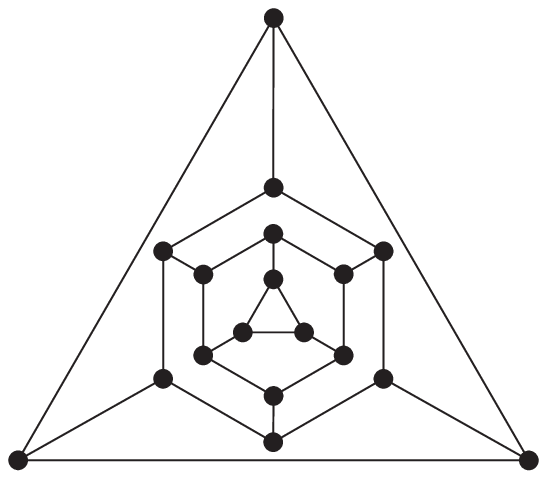}&
\includegraphics[scale=0.65]{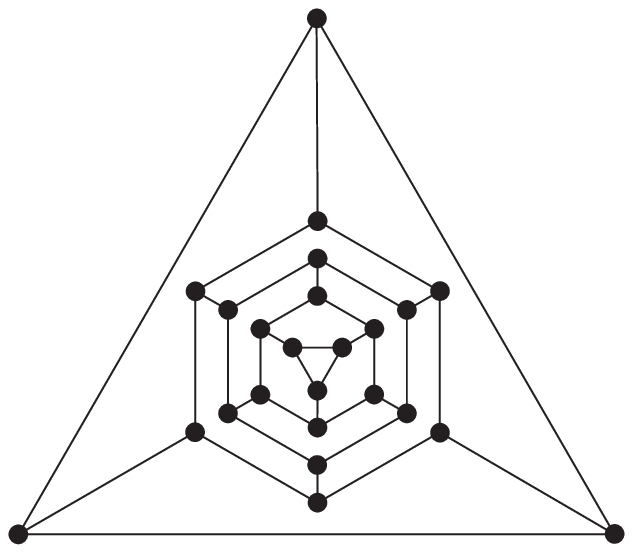}\\
{\small (a)}&{\small (b)}
\end{tabular}
\caption{The graphs that result from the second (a) and third (b) substitutions
of $I_3$ for $S_3$.}
\vspace{0.25in}
\label{2iter&3iter}
\end{figure}

Notice that we can now replace $S_3$ by $I_3$ in $B_3$, thus obtaining a larger
cubic graph of girth $3$ (the one in Figure~\ref{I3&1iter}(b)) that has a 
$6$-CDC. This substitution process can proceed recursively, always generating 
cubic graphs of girth $3$ with a $6$-CDC. The graphs resulting from the second 
and third iterations are shown in Figure~\ref{2iter&3iter}. 

\subsection{The girth-$\mathbf{4}$ case}
\label{girth4}

\begin{figure}[t]
\centering
\begin{tabular}{c@{\hspace{2.0cm}}c@{\hspace{2.0cm}}c}
\includegraphics[scale=0.65]{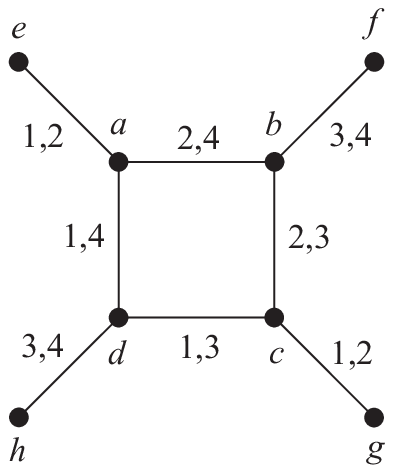}&
\includegraphics[scale=0.65]{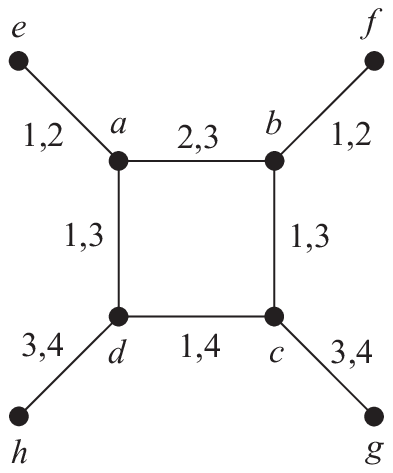}&
\includegraphics[scale=0.65]{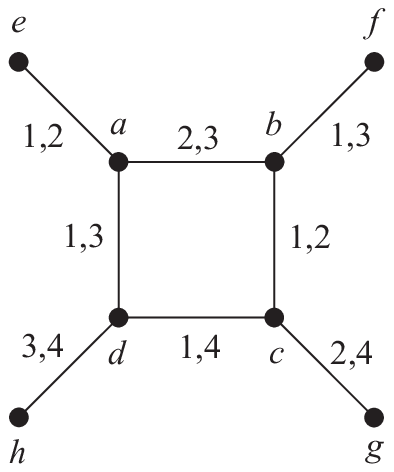}\\
{\small (a)}&{\small (b)}&{\small (c)}
\end{tabular}
\caption{$S_{4\mathrm{a}}$ (a), $S_{4\mathrm{b}}$ (b), and $S_{4\mathrm{c}}$ (c).}
\vspace{0.25in}
\label{s4}
\end{figure}

We start by analyzing all the possible cycle configurations of $S_4$. In 
Figures~\ref{s4}(a)--(c), the cycle configurations of 
$S_4$ that infringe neither Lemma~\ref{nopath} nor Lemma~\ref{3cycles}, and also 
do not lead to the existence of a cycle with length smaller than $6$ in the 
$6$-CDC, are presented. Note that vertices $e$, $f$, $g$, and $h$ are all 
distinct in these graphs. The cases in which these vertices may coincide will be 
treated later.

For each cycle configuration shown in Figure~\ref{s4}, we must identify $I_4$ 
and $B_4$. We denote by $S_{4\mathrm{a}}$ the graph with the cycle configuration 
of Figure~\ref{s4}(a), and likewise $I_{4\mathrm{a}}$ and $B_{4\mathrm{a}}$ 
refer to the expansions of $S_{4\mathrm{a}}$. We proceed similarly in the cases 
of Figures~\ref{s4}(b) and \ref{s4}(c).\footnote{The expansion of 
$S_{4\mathrm{b}}$ into a $B_4$ instance has two possible outcomes, which we denote 
by $B_{4\mathrm{b}}$ and $B_{4\mathrm{b'}}$.}

\begin{figure}[p]
\centering
\begin{tabular}{c@{\hspace{1.5cm}}c}
\includegraphics[scale=0.65]{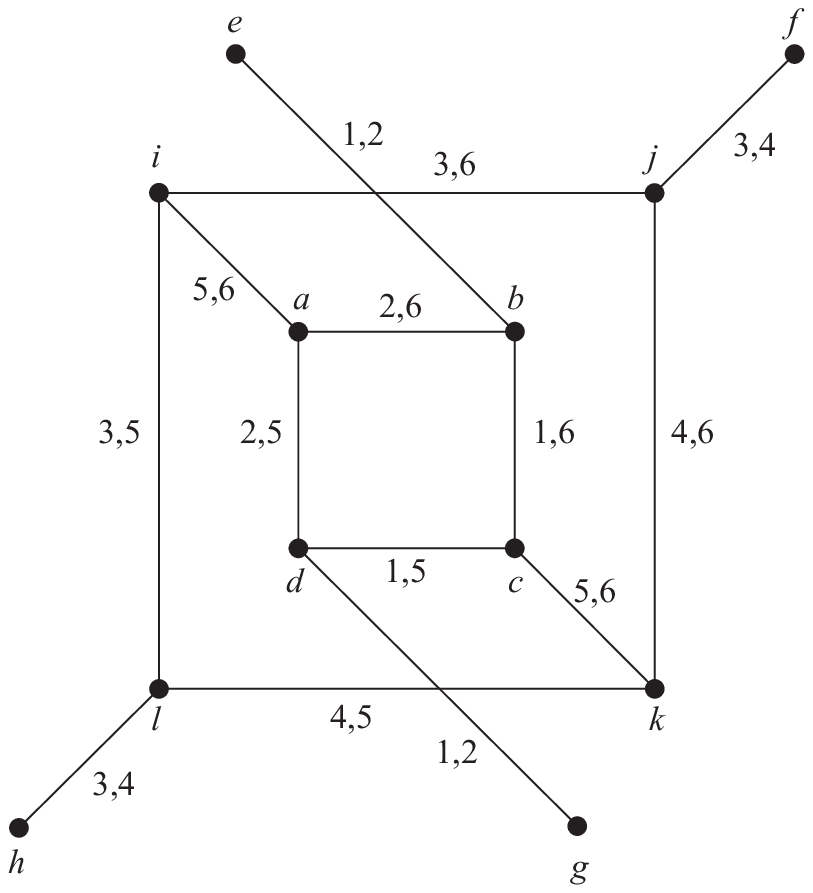}&
\includegraphics[scale=0.65]{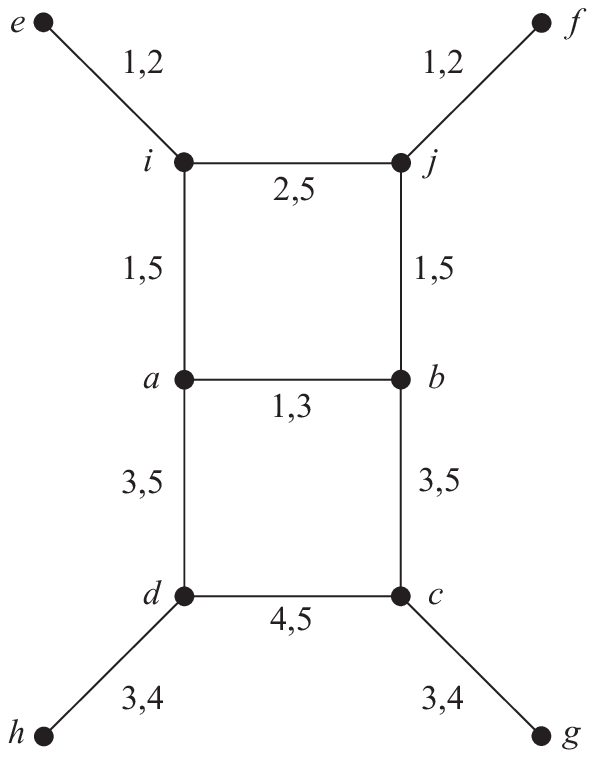}\\
{\small (a)}&{\small (b)}
\vspace{0.5cm}
\end{tabular}
\begin{tabular}{c}
\includegraphics[scale=0.65]{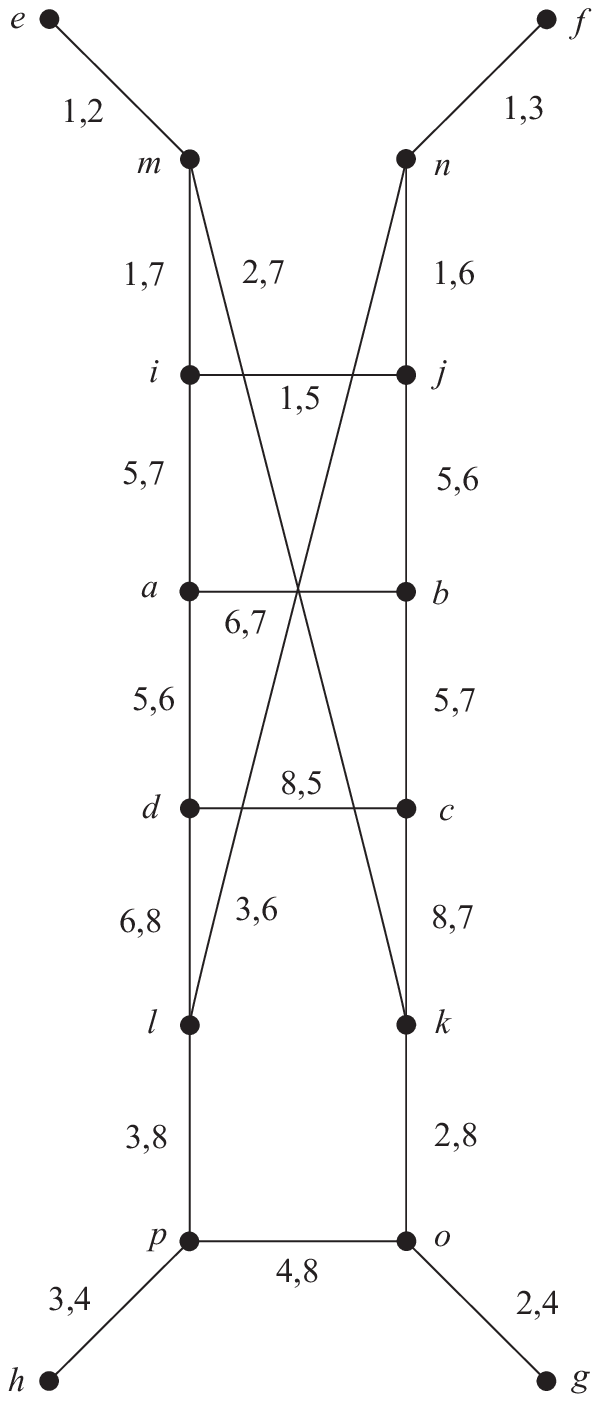}\\
{\small (c)}
\end{tabular}
\caption{$I_{4\mathrm{a}}$ (a), $I_{4\mathrm{b}}$ (b), and $I_{4\mathrm{c}}$ 
(c).}
\vspace{0.25in}
\label{i4}
\end{figure}

\begin{figure}[p]
\centering
\begin{tabular}{c@{\hspace{1.5cm}}c}
\includegraphics[scale=0.65]{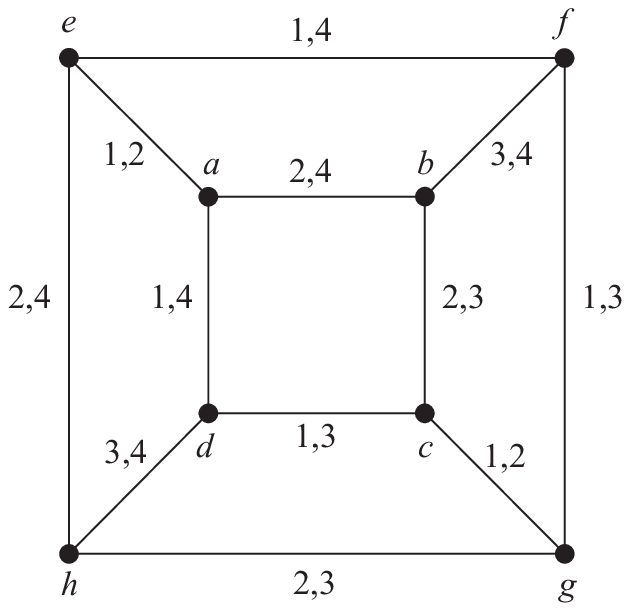}&
\includegraphics[scale=0.65]{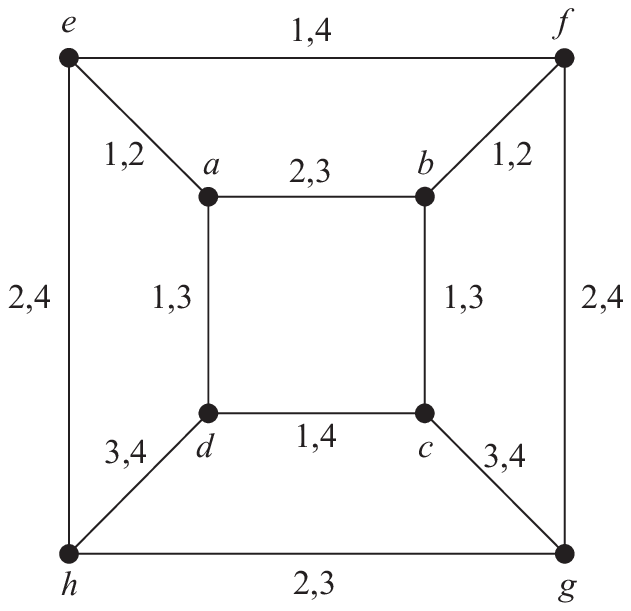}\\
{\small (a)}&{\small (b)}
\vspace{0.5cm}
\end{tabular}
\begin{tabular}{c@{\hspace{1.5cm}}c}
\includegraphics[scale=0.65]{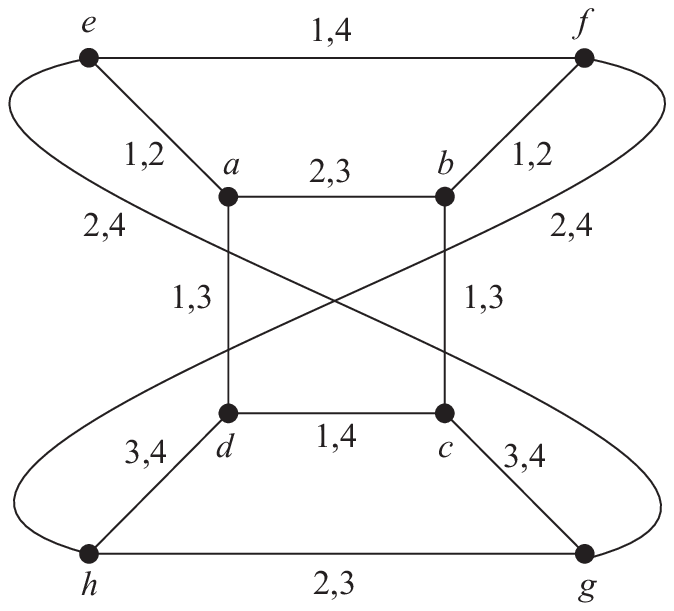}&
\includegraphics[scale=0.65]{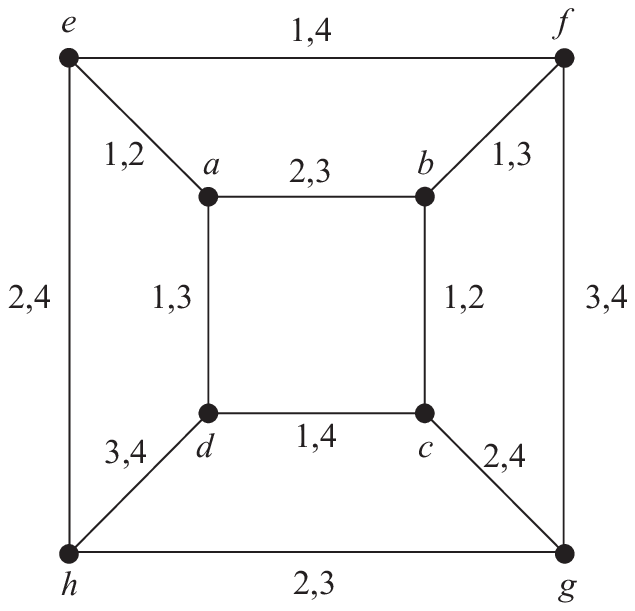}\\
{\small (c)}&{\small (d)}
\end{tabular}
\caption{$B_{4\mathrm{a}}$ (a), $B_{4\mathrm{b}}$ (b), $B_{4\mathrm{b'}}$ (c), 
and $B_{4\mathrm{c}}$ (d).}
\vspace{0.25in}
\label{b4}
\end{figure}

The only possibilities for $I_4$ are the $I_{4\mathrm{a}}$, 
$I_{4\mathrm{b}}$, and $I_{4\mathrm{c}}$ of Figure~\ref{i4}.
As for $B_4$, the only possibilities that Lemmas~\ref{nopath} and \ref{3cycles} 
allow are the $B_{4\mathrm{a}}$, $B_{4\mathrm{b}}$, $B_{4\mathrm{b'}}$, and 
$B_{4\mathrm{c}}$ of Figure~\ref{b4}.
Notice that each of $B_{4\mathrm{a}}$, $B_{4\mathrm{b}}$, and $B_{4\mathrm{c}}$ is 
isomorphic to $T_{8,2}$, and that $B_{4\mathrm{b'}}$ is isomorphic to $M_8$.
Also, they are all type-(i) instances of $B_4$.

It is illustrative to notice also that the cycle configurations of $S_{4\mathrm{a}}$
and $I_{4\mathrm{a}}$ are in fact equivalent to each other, and also that
every subgraph of $I_{4\mathrm{a}}$ that is isomorphic to $S_{4\mathrm{a}}$ has 
a cycle configuration that is equivalent to that of $S_{4\mathrm{a}}$. That is,
$I_{4\mathrm{a}}$ is indeed self-similar with respect to $S_{4\mathrm{a}}$ and does 
as such allow for recursive substitutions of $I_{4\mathrm{a}}$ for 
$S_{4\mathrm{a}}$ starting at $B_{4\mathrm{a}}$. Except for the initial 
$B_{4\mathrm{a}}$, since it is of type (i),
all $S_{4\mathrm{a}}$-isomorphic subgraphs of the resulting 
graphs have cycle configurations equivalent to that of
$S_{4\mathrm{a}}$.\footnote{In Section~\ref{complet},
we use this property to argue for the 
uniqueness of each graph generated in the process.} The cases of 
$S_{4\mathrm{b}}$ and $I_{4\mathrm{b}}$ with $B_{4\mathrm{b}}$ (or 
$B_{4\mathrm{b'}}$) and of $S_{4\mathrm{c}}$ and $I_{4\mathrm{c}}$ with 
$B_{4\mathrm{c}}$ are entirely analogous.

Now let us consider the cases in which vertices $e$, $f$, $g$, and $h$ are not 
necessarily distinct. It is easy to see that the only way for this to happen 
without altering the girth is to let $e=g$ or $f=h$. If either $e=g$ or $f=h$, 
then clearly the cycle configurations of Figures~\ref{s4}(a) and \ref{s4}(c) 
acquire a cycle of length $4$, which is inconsistent with the nature of a 
$6$-CDC, while in the cycle configuration of Figure~\ref{s4}(b) either vertex 
$e$ or vertex $f$ becomes part of four distinct cycles, which infringes
Lemma~\ref{3cycles}. 

\begin{figure}[t]
\centering
\includegraphics[scale=0.65]{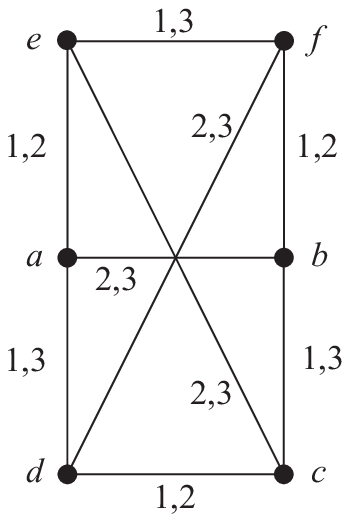}
\caption{Graph, with a $6$-CDC, obtained from $S_{4\mathrm{b}}$ by letting 
$e=g$, $f=h$, and $C_2=C_4$ while $e$ is connected to $f$.}
\vspace{0.25in}
\label{m6}
\end{figure}

Letting both $e=g$ and $f=h$, similarly, violates the $6$-CDC in the cases of 
Figures~\ref{s4}(a) and \ref{s4}(c). However, the cycle configuration of 
Figure~\ref{s4}(b) remains valid, and by simply adding edge $ef$ and letting 
$C_2=C_4$ we obtain $M_6$ with a consistent $6$-CDC, as in Figure~\ref{m6}. It 
is interesting to note that the first replacement of $S_{4\mathrm{b}}$ by 
$I_{4\mathrm{b}}$ in $M_6$ generates $M_8$, which is isomorphic to 
$B_{4\mathrm{b'}}$, so we may actually let $B_{4\mathrm{b'}}$ be $M_6$ instead
(and thus avoid creating another type-(i) instance of $B_4$). 

\subsection{The girth-$\mathbf{5}$ case}
\label{girth5}

\begin{figure}[p]
\centering
\begin{tabular}{c@{\hspace{1.5cm}}c}
\includegraphics[scale=0.65]{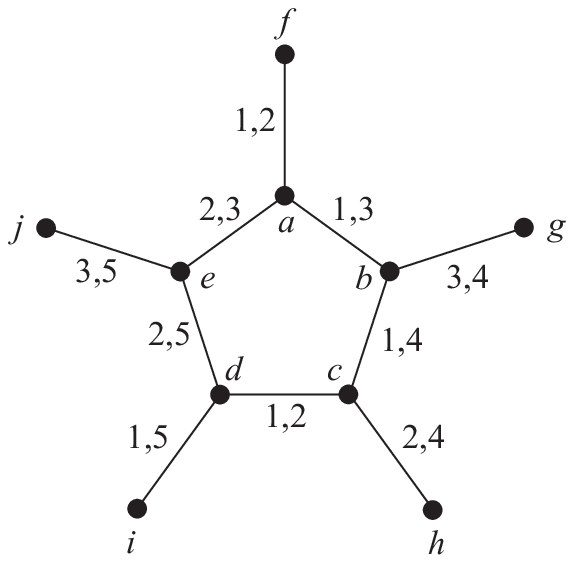}&
\includegraphics[scale=0.65]{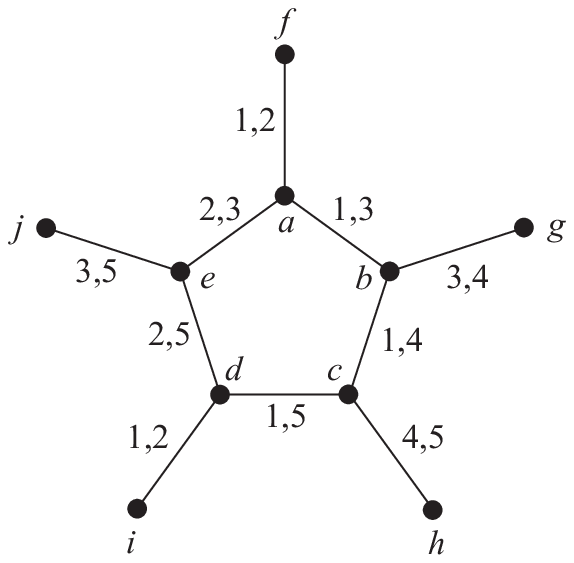}\\
{\small (a)}&{\small (b)}
\vspace{0.5cm}
\end{tabular}
\begin{tabular}{c}
\includegraphics[scale=0.65]{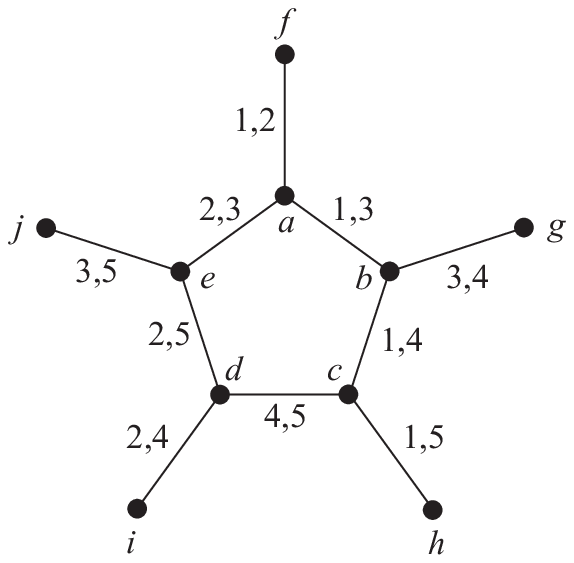}\\
{\small (c)}
\end{tabular}
\caption{$S_{5\mathrm{a}}$ (a), $S_{5\mathrm{b}}$ (b), and $S_{5\mathrm{c}}$ 
(c).}
\vspace{0.25in}
\label{s5}
\end{figure}

In Figures~\ref{s5}(a)--(c), the cycle configurations 
of $S_5$ that are consistent with Lemmas~\ref{nopath} and \ref{3cycles} and 
do not disrupt the nature of the $6$-CDC are depicted. Notice that vertices $f$, 
$g$, $h$, $i$, and $j$ must necessarily be distinct in order for the girth 
not to fall below $5$. As in the girth-$4$ case, these graphs along with 
their cycle configurations are denoted by $S_{5\mathrm{a}}$, $S_{5\mathrm{b}}$, 
and $S_{5\mathrm{c}}$, respectively.

\begin{figure}[t]
\centering
\includegraphics[scale=0.65]{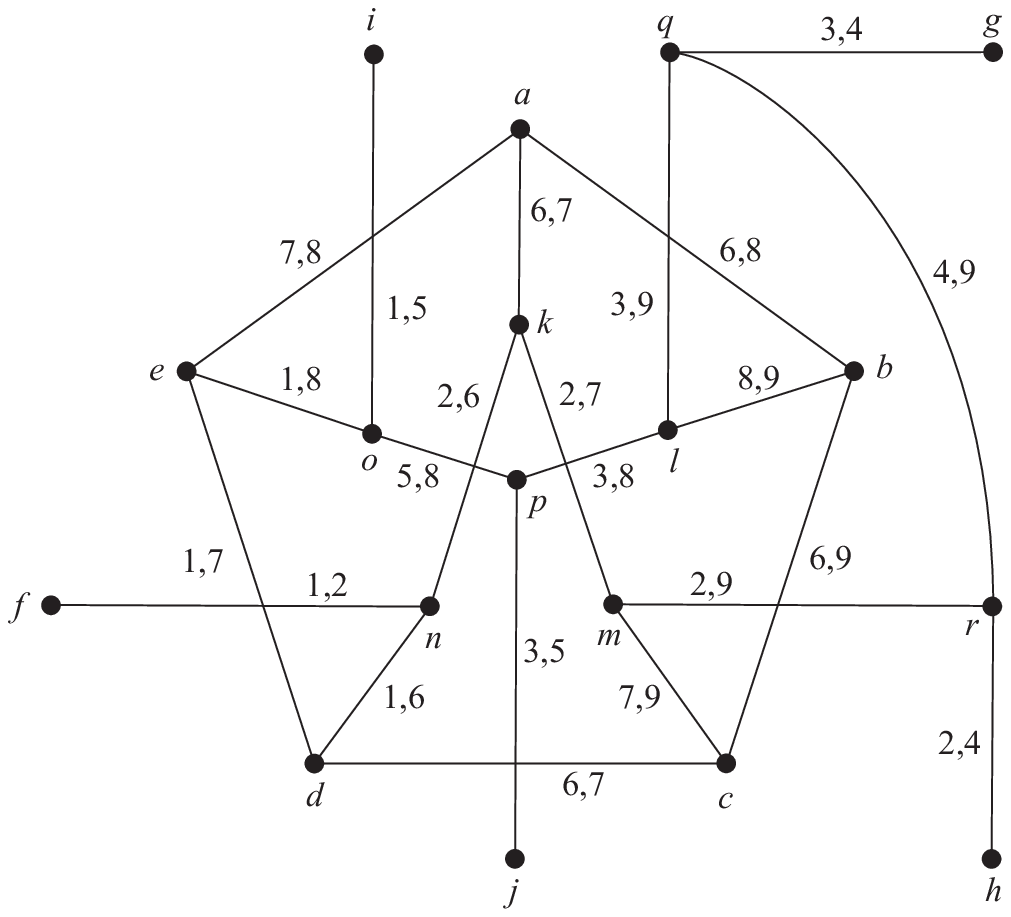}
\caption{$I_{5\mathrm{a}}$.}
\vspace{0.25in}
\label{i5a}
\end{figure}

Now, as we try to expand $S_{5\mathrm{a}}$, we invariably generate the
$I_{5\mathrm{a}}$ of Figure~\ref{i5a} before we get to a $B_5$ instance, and do
so without ever turning down an edge addition exclusively on account that the
graph's girth would be thus reduced. One consequence of this is that any $B_5$
instance we may come to generate as we proceed with the expansion will have
$I_{5\mathrm{a}}$ as a subgraph and therefore not be of type (i). Furthermore,
as we consider that the cycle configurations of $S_{5\mathrm{a}}$ and
$I_{5\mathrm{a}}$ are equivalent to each other, we realize that expanding
beyond $I_{5\mathrm{a}}$ is almost completely constrained to repeating the same
steps that initially led from $S_{5\mathrm{a}}$ to $I_{5\mathrm{a}}$. The only
exception is that now we may be precluded from adding a certain edge solely
because such an addition would reduce the graph's girth.\footnote{The fact that 
adding an edge $uv$ to $I_g$ creates a cycle with length smaller than $g$ does 
not necessarily imply that an edge $u'v'$ in $S_g$ will form a cycle with length
smaller than $g$ as well, where $u'$ and $v'$ in $I_g$ correspond to $u$ and $v$
in $S_g$, respectively.} But since nothing of this sort happens in the expansion
from $S_{5\mathrm{a}}$ to $I_{5\mathrm{a}}$, we see in any event that
substituting $S_{5\mathrm{a}}$ for $I_{5\mathrm{a}}$ in any deficiency-$0$ graph
resulting from expanding beyond $I_{5\mathrm{a}}$ preserves the girth, and then
that graph is not a type-(ii) $B_5$ instance. So it turns out that no $B_5$
instance can be generated, and then the cycle configuration of $S_{5\mathrm{a}}$
is invalid. As for $S_{5\mathrm{b}}$, it is relatively easy to see that its
expansion cannot proceed without infringing Lemma~\ref{nopath} or reducing the
graph's girth. This cycle configuration is therefore also invalid.

\begin{figure}[p]
\centering
\includegraphics[scale=0.65]{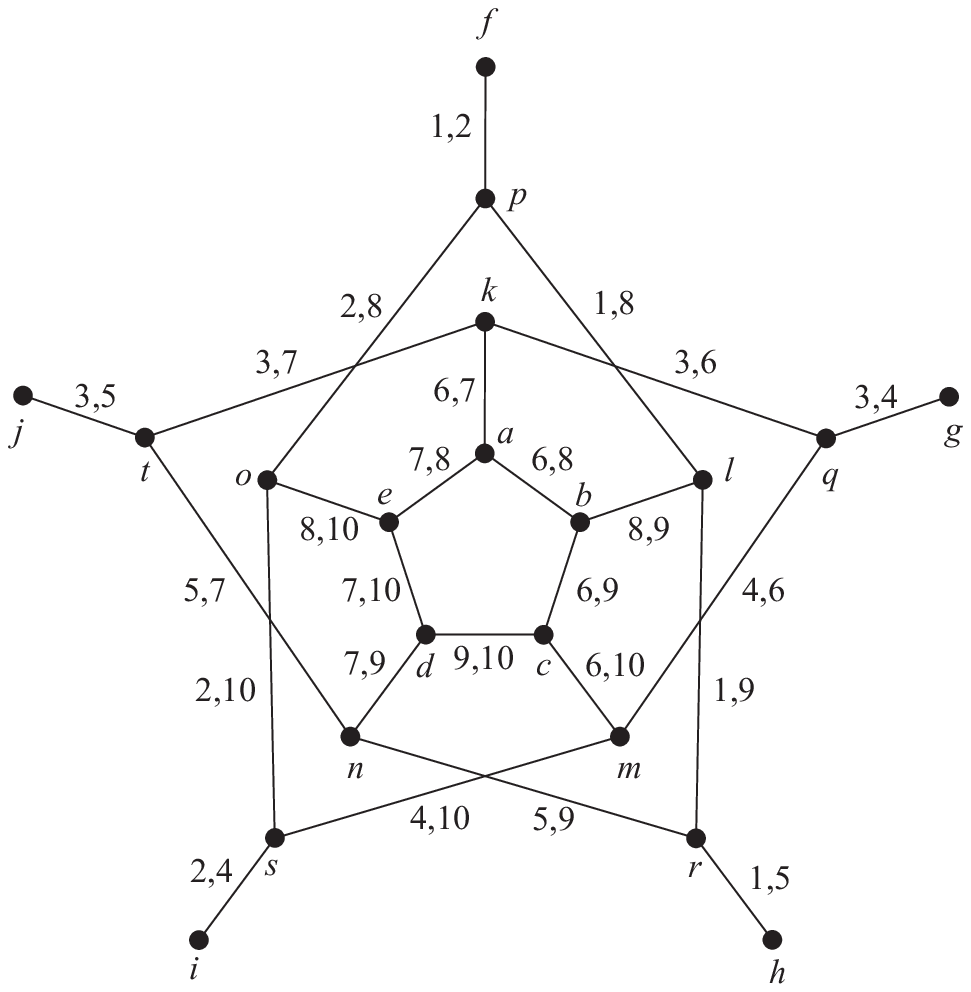}
\caption{$I_{5\mathrm{c}}$.}
\vspace{0.25in}
\label{i5c}
\end{figure}

\begin{figure}[p]
\centering
\includegraphics[scale=0.65]{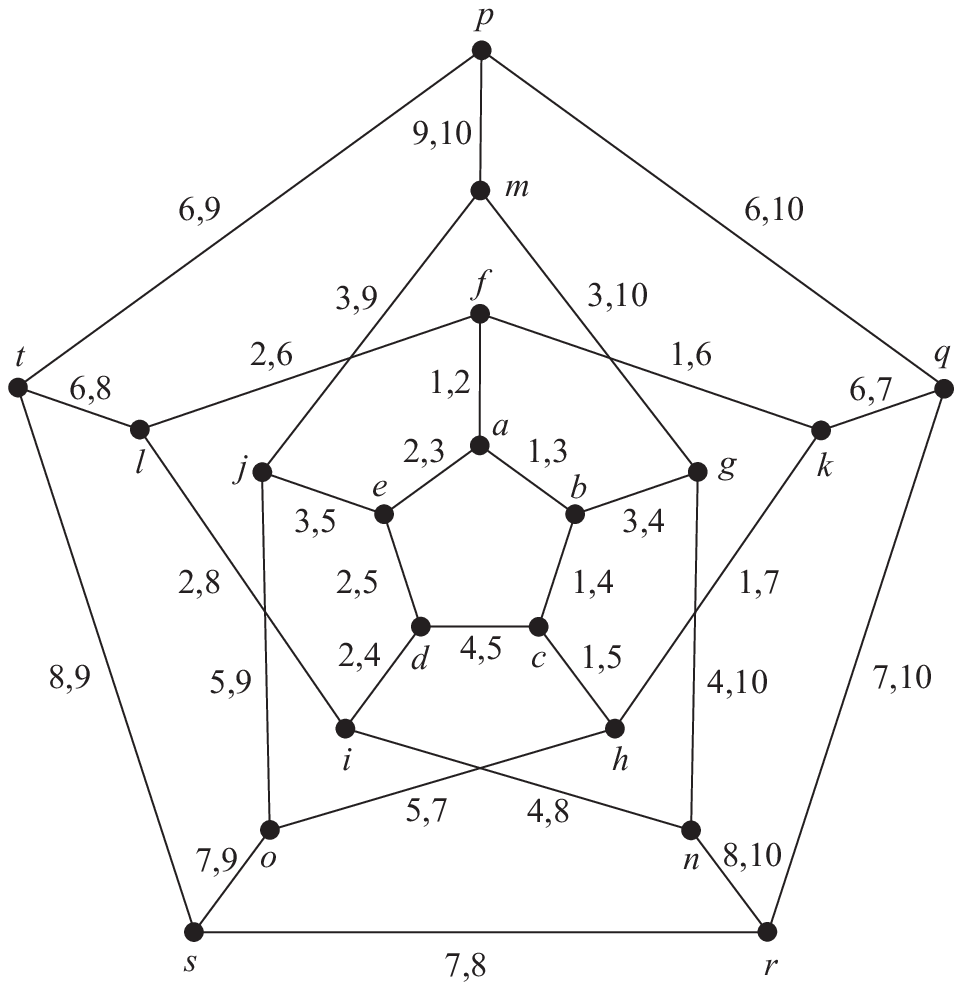}
\caption{$B_{5\mathrm{c}}$.}
\vspace{0.25in}
\label{b5c}
\end{figure}

The expansion of $S_{5\mathrm{c}}$, on the other hand, leads to the 
$I_{5\mathrm{c}}$ of Figure~\ref{i5c}. And even though we arrive at 
$I_{5\mathrm{c}}$ before obtaining $B_{5\mathrm{c}}$, this expansion does
refrain from adding edges that would reduce the graph's girth.
So we may proceed with the expansion of $I_{5\mathrm{c}}$ 
until we generate the $B_{5\mathrm{c}}$ of Figure~\ref{b5c}, which
is a type-(ii) instance of $B_5$. As before, it is 
important to note that the two subgraphs of $B_{5\mathrm{c}}$ isomorphic to 
$S_{5\mathrm{c}}$ have the same cycle configuration as $S_{5\mathrm{c}}$. 
   
\subsection{The girth-$\mathbf{6}$ case}
\label{girth6}

\begin{figure}[p]
\centering
\begin{tabular}{c@{\hspace{1.5cm}}c}
\includegraphics[scale=0.65]{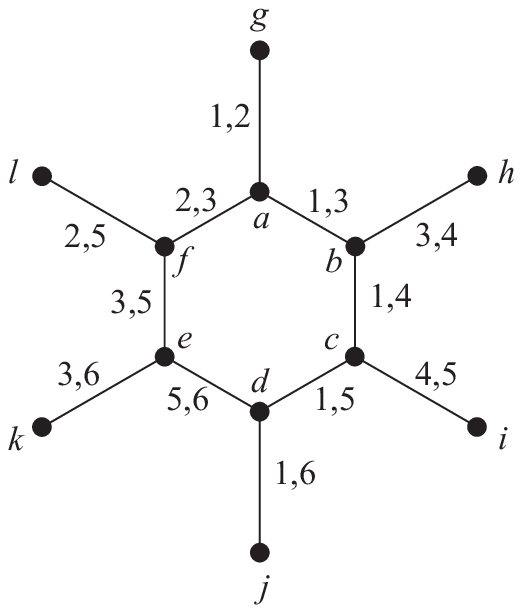}&
\includegraphics[scale=0.65]{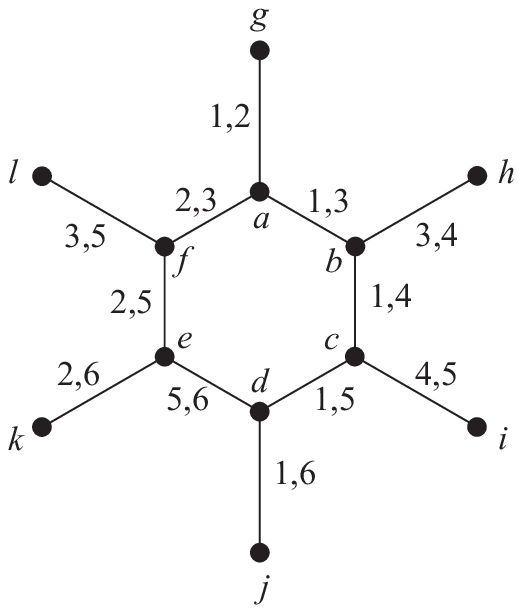}\\
{\small (a)}&{\small (b)}
\vspace{0.5cm}
\end{tabular}
\begin{tabular}{c@{\hspace{1.5cm}}c}
\includegraphics[scale=0.65]{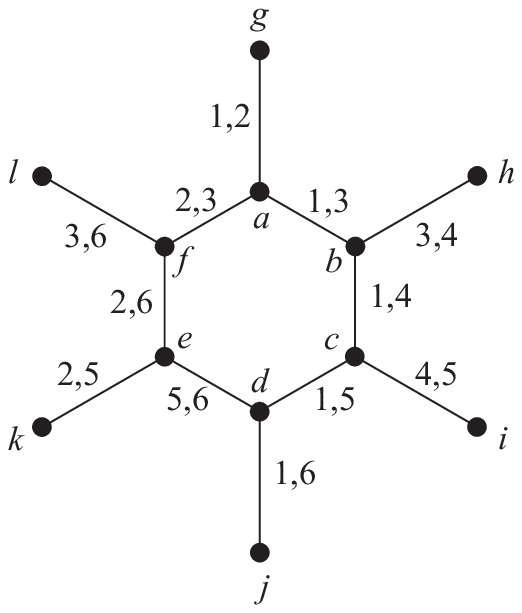}&
\includegraphics[scale=0.65]{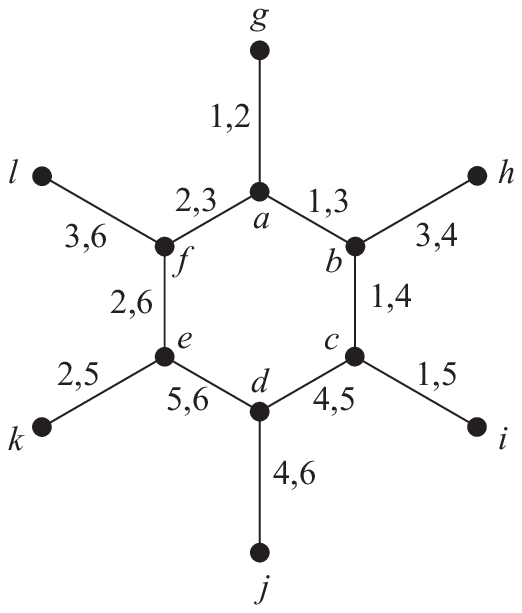}\\
{\small (c)}&{\small (d)}
\vspace{0.5cm}
\end{tabular}
\begin{tabular}{c}
\includegraphics[scale=0.65]{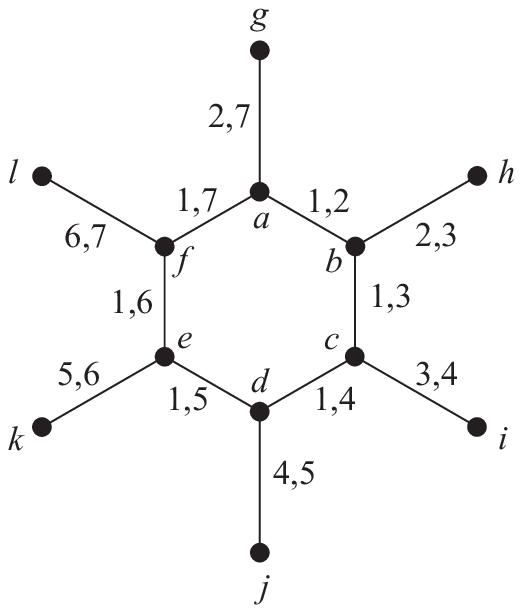}\\
{\small (e)}
\end{tabular}
\caption{$S_{6\mathrm{a}}$ (a), $S_{6\mathrm{b}}$ (b), $S_{6\mathrm{c}}$ (c), 
$S_{6\mathrm{d}}$ (d), and $S_{6\mathrm{e}}$ (e).}
\vspace{0.25in}
\label{s6}
\end{figure}

\begin{figure}[p]
\centering
\begin{tabular}{c@{\hspace{1.5cm}}c}
\includegraphics[scale=0.65]{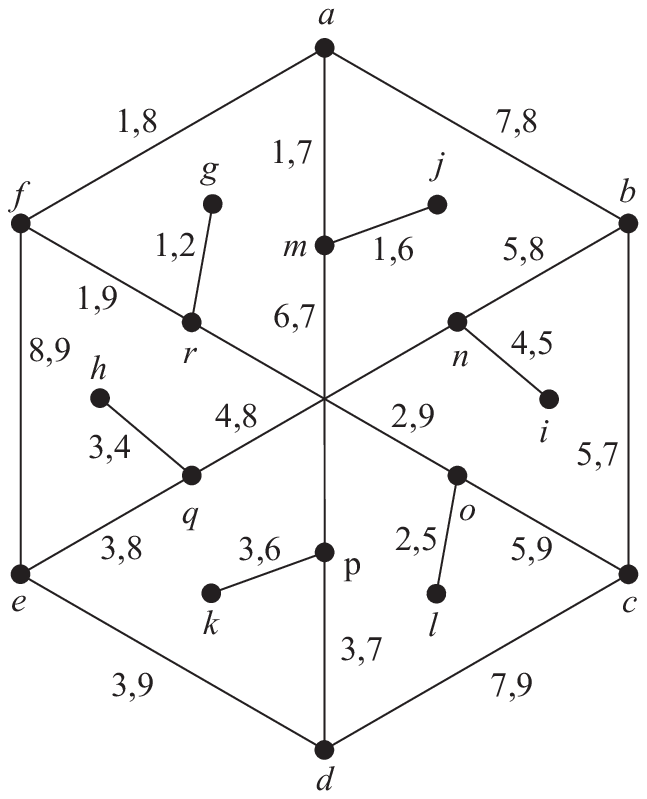}&
\includegraphics[scale=0.65]{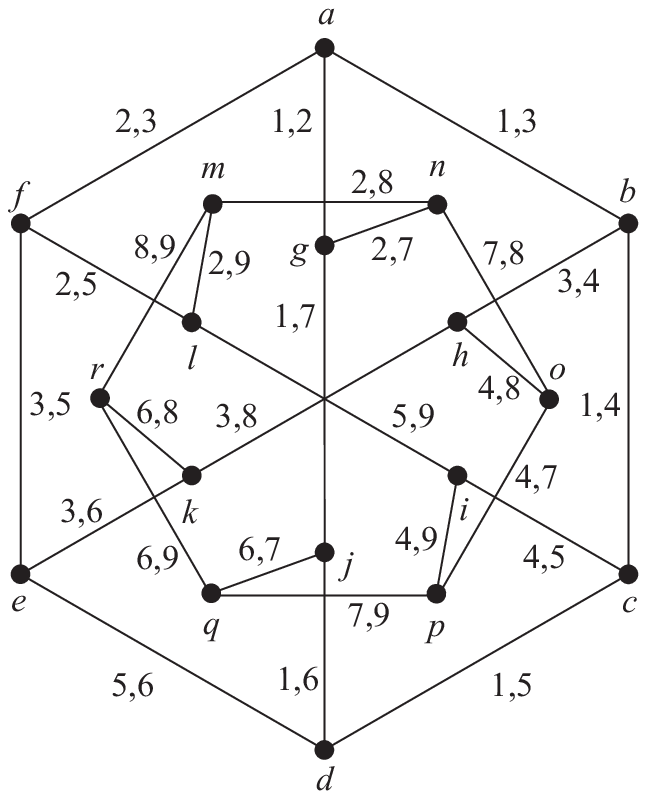}\\
{\small (a)}&{\small (b)}
\end{tabular}
\begin{tabular}{c}
\includegraphics[scale=0.65]{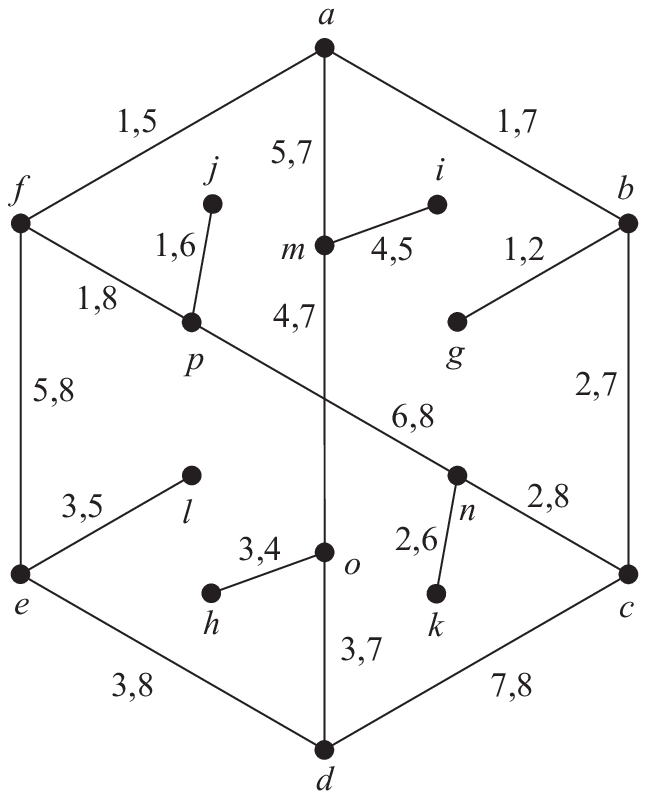}\\
{\small (c)}
\end{tabular}
\begin{tabular}{c@{\hspace{1.5cm}}c}
\includegraphics[scale=0.65]{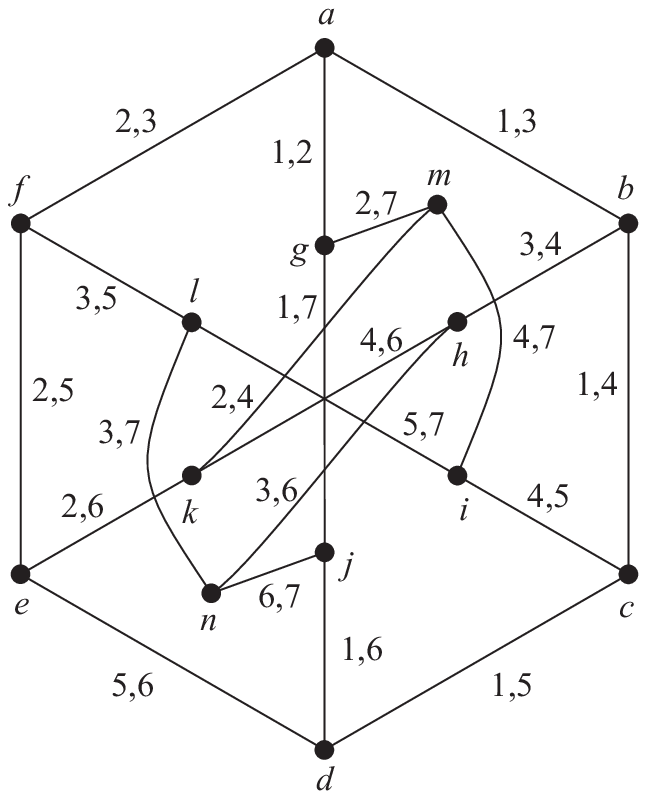}&
\includegraphics[scale=0.65]{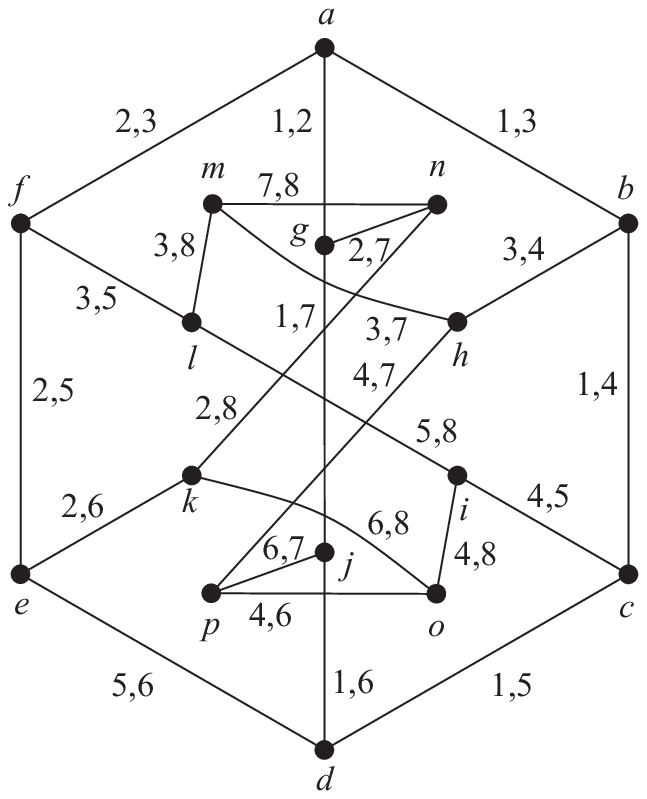}\\
{\small (d)}&{\small (e)}
\end{tabular}
\caption{$I_{6\mathrm{a}}$ (a), $B_{6\mathrm{a}}$ (b), $I_{6\mathrm{b}}$ (c), 
$B_{6\mathrm{b}}$ (d), $B_{6\mathrm{b'}}$ (e).}
\label{bi6ab}
\end{figure}

\begin{figure}[p]
\centering
\begin{tabular}{c@{\hspace{1.5cm}}c}
\includegraphics[scale=0.65]{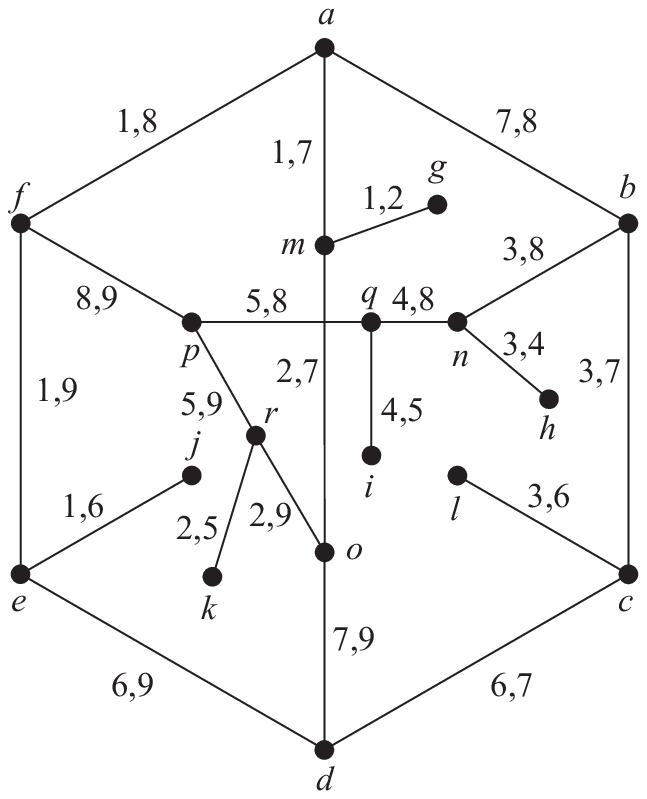}&
\includegraphics[scale=0.65]{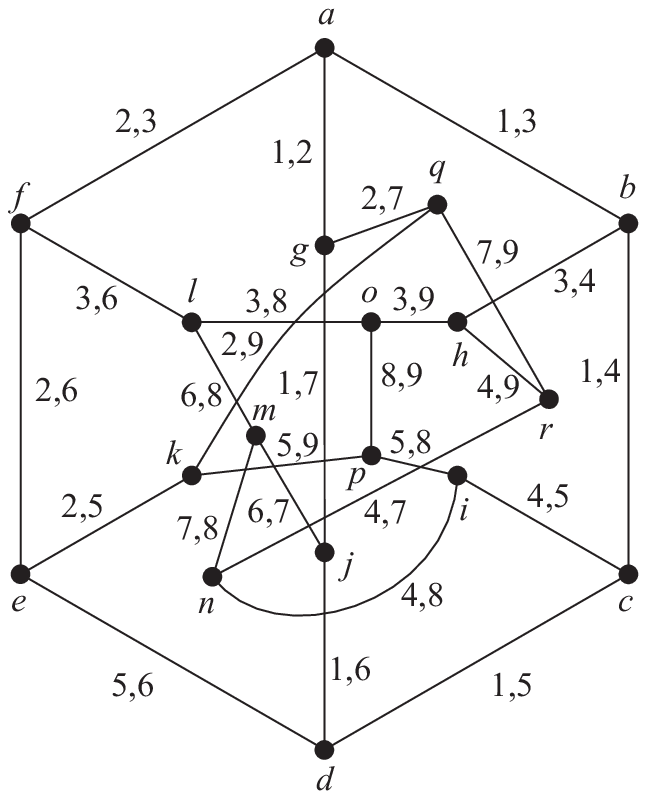}\\
{\small (a)}&{\small (b)}
\end{tabular}
\begin{tabular}{c}
\includegraphics[scale=0.65]{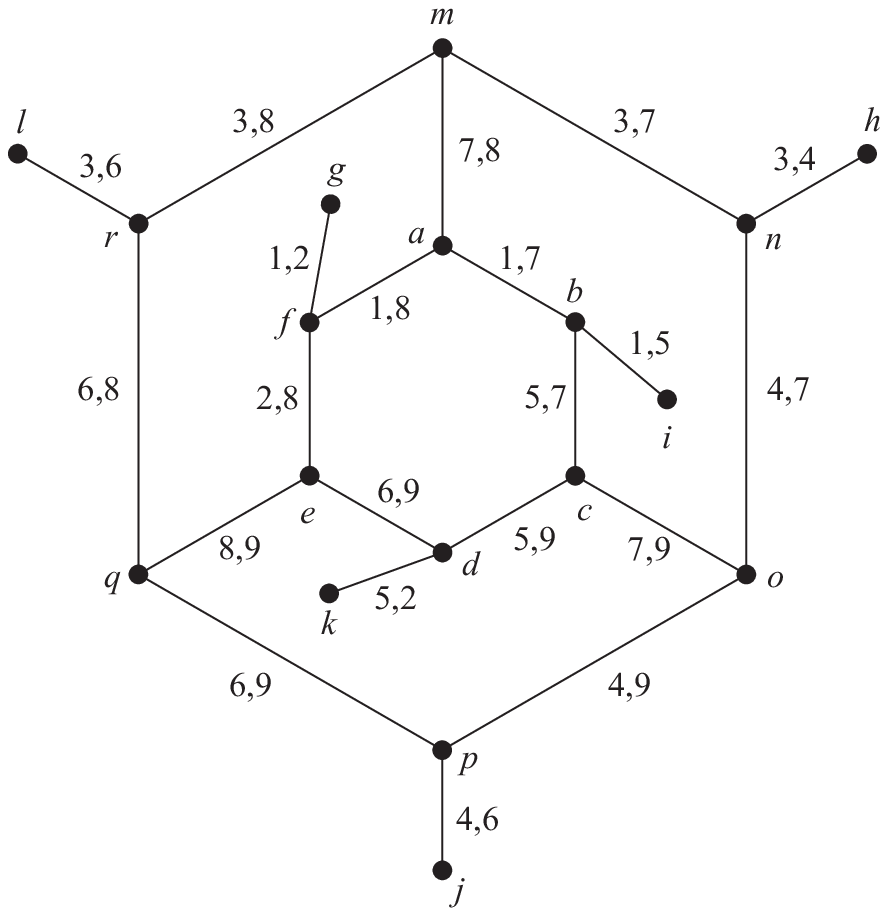}\\
{\small (c)}
\vspace{-0.75cm}
\end{tabular}
\begin{tabular}{c@{\hspace{1.5cm}}c}
\includegraphics[scale=0.65]{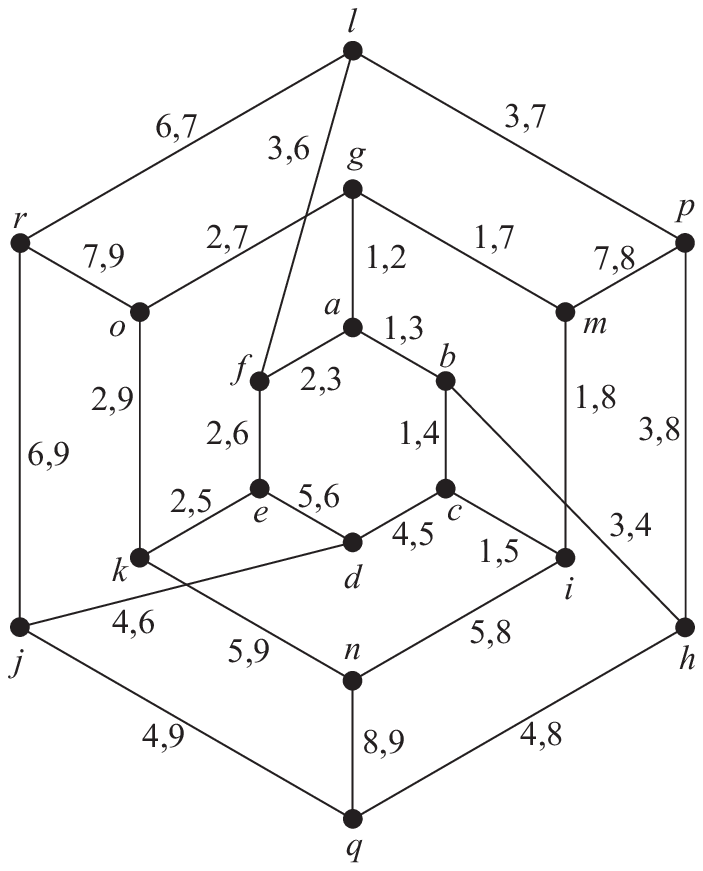}&
\includegraphics[scale=0.65]{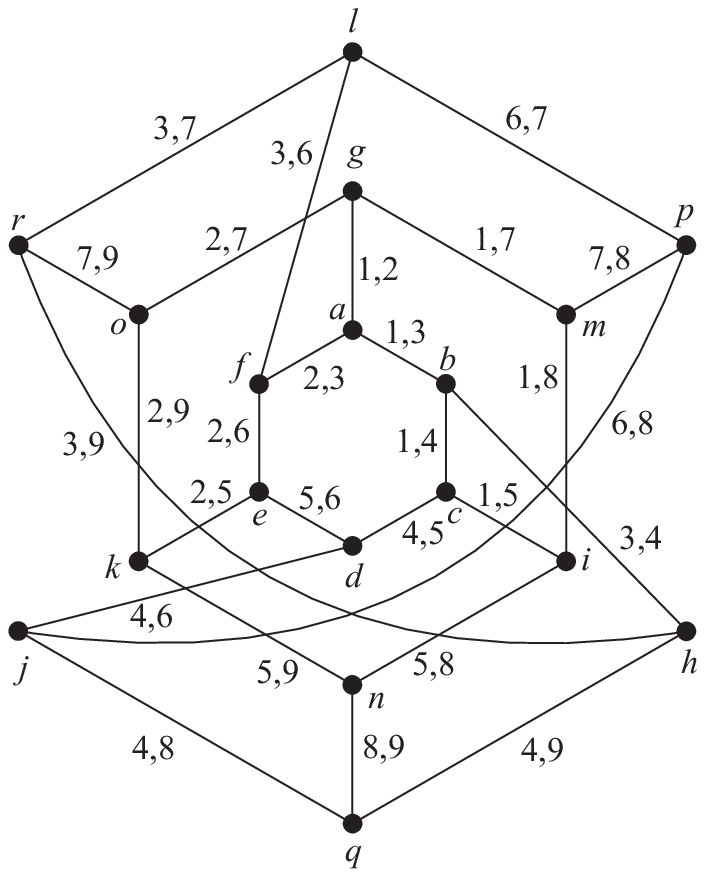}\\
{\small (d)}&{\small (e)}
\end{tabular}
\caption{$I_{6\mathrm{c}}$ (a), $B_{6\mathrm{c}}$ (b), $I_{6\mathrm{d}}$ (c),  
$B_{6\mathrm{d}}$ (d), and $B_{6\mathrm{d'}}$ (e).}
\label{bi6cd}
\end{figure}

Following our development so far, we present in Figures~\ref{s6}(a)--(e) the 
consistent cycle configurations of $S_6$, namely $S_{6\mathrm{a}}$ through 
$S_{6\mathrm{e}}$. Of these, $S_{6\mathrm{e}}$ is the only one to include a
complete $6$-CDC cycle (cycle $C_1$) in its cycle configuration.
The $I_6$ and $B_6$ instances for $S_{6\mathrm{a}},\ldots,S_{6\mathrm{d}}$ are given
in Figures~\ref{bi6ab} and \ref{bi6cd}. Notice that, consistently with our comments
in Section~\ref{method}, every one of $I_{6\mathrm{a}}$ through $I_{6\mathrm{d}}$ has
subgraphs that are isomorphic to $S_6$ but do not have
the same cycle configuration as, respectively, $S_{6\mathrm{a}}$ through $S_{6\mathrm{d}}$
(having, as those subgraphs do, a complete $6$-CDC cycle in their
cycle configurations). Notice also that $B_{6\mathrm{b}}$ (the Heawood graph \cite{bondy})
is the only type-(i) instance of $B_6$ in the group; the others are all of type (ii).
The case of $S_{6\mathrm{e}}$, however, 
embodies peculiarities we have not yet encountered, and does as such require 
further elaboration.

As we noted above, $S_{6\mathrm{e}}$ contains a complete cycle from the $6$-CDC ($C_1$ 
in Figure~\ref{s6}(e)). Also, it represents the only possible cycle configuration for a 
$6$-CDC cycle in a girth-$6$ graph. It then follows that $S_{6\mathrm{e}}$ is 
contained in every girth-$6$ graph that has a $6$-CDC. However, there exist 
graphs that have a $6$-CDC and $S_{6\mathrm{e}}$ as a subgraph (including its 
cycle configuration) but do not have $S_{6\mathrm{a}}$, $S_{6\mathrm{b}}$, 
$S_{6\mathrm{c}}$, or $S_{6\mathrm{d}}$ as subgraphs: they are the graphs in 
which every $6$-cycle belongs to the $6$-CDC. So, analogously to our strategy 
thus far, let us look for corresponding $I_6$ and $B_6$ instances. From
Section~\ref{method}, we know that the desired $B_6$ instances are of type (iii).

In order to facilitate the task of searching for $I_{6\mathrm{e}}$
and also for $B_{6\mathrm{e}}$, we first 
look into some properties related to the graph's girth.

\begin{lem}
\label{cycle_adjacency}
If $C$ is a cycle in a $6$-CDC of $G$ such that $\sigma(C)<6$, then $g(G)<5$.
\end{lem}
\begin{proof}
If $C$ has a chord, then the lemma holds trivially. Let us then assume that 
$C$ is chordless. In this case, there has to exist another $6$-CDC cycle, say 
$C'$, such that $\mu(C, C')>1$, since $\sigma(C)<6$. By Lemma~\ref{boundmu}, 
$\mu(C, C')\leq 3$; by Lemma~\ref{mu3}, if $\mu(C,C')=3$ then $G$ is isomorphic 
to either $M_6$ or $T_{6,2}$ and, consequently, every $6$-CDC cycle has a chord, 
which cannot be by assumption. Thus, $\mu(C,C')=2$.

Let $e$ and $f$ be edges of $G$ belonging to both $C$ and $C'$. Because $C$ is 
chordless, $C'$ contains two distinct paths, call them $P_1$ and $P_2$, of 
length $2$, both interconnecting an end vertex of $e$ with an end vertex of $f$, 
as in Figure~\ref{paths}. Let $u$ and $v$ be end vertices of $e$ and $f$, 
respectively, such that $P_1$ interconnects $u$ and $v$. It is easy to see that 
there exists another path in $C$, call it $P_3$, that interconnects $u$ and $v$ 
in such a way that $P_3$ has length less than $4$. Hence, there exists a cycle 
in $G$ containing $u$ and $v$ whose length is less than $6$ and whose edges are 
those of $P_1$ and $P_3$. 

\begin{figure}[t]
\centering
\includegraphics[scale=0.65]{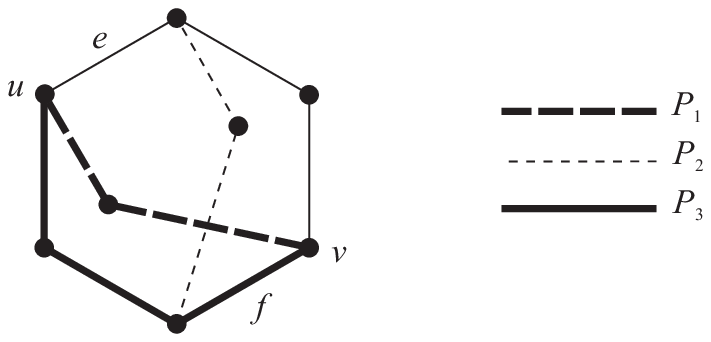}
\caption{Subgraph induced by $C$ and $C'$. Dashed edges belong to $C'$.}
\vspace{0.25in}
\label{paths}
\end{figure}

However, there is only one way, shown in Figure~\ref{paths}, of interconnecting 
the end vertices of $e$ and $f$ such that the subgraph of $G$ induced by the 
edges of $C$ and $C'$ has girth $5$. And there exists only one possible cycle 
configuration for this subgraph, considering the already given cycles $C$ and 
$C'$. This cycle configuration contains $S_{5\mathrm{a}}$, which from Section~\ref{girth5}
we know is not valid. So the interconnection pattern of 
Figure~\ref{paths} is invalid as well, thence $g(G)<5$.             
\end{proof}

\begin{thm}
\label{girth}
$g(G) \geq 5$ if and only if every cycle $C$ in a $6$-CDC of $G$ is such that $\sigma(C)=6$.
\end{thm}
\begin{proof}
If $\sigma(C)<6$ for some $6$-CDC cycle $C$, then $g(G)<5$ by Lemma~\ref{cycle_adjacency}.
In order to prove the converse statement, we assume 
$g(G)=3$ or $g(G)=4$ and simply verify, respectively from Figure~\ref{b3&s3}(b) 
or Figures~\ref{s4}(a)--(c), that $6$-CDC cycles $C$ and $C'$ always exist such 
that $\mu(C,C')>1$. It then follows that a $6$-CDC cycle $C$ always exists for 
which $\sigma(C)<6$.
\end{proof}

\begin{figure}[t]
\centering
\begin{tabular}{c@{\hspace{1.5cm}}c}
\includegraphics[scale=0.65]{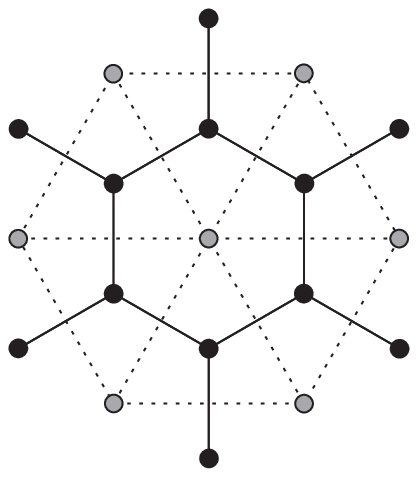}&
\includegraphics[scale=0.65]{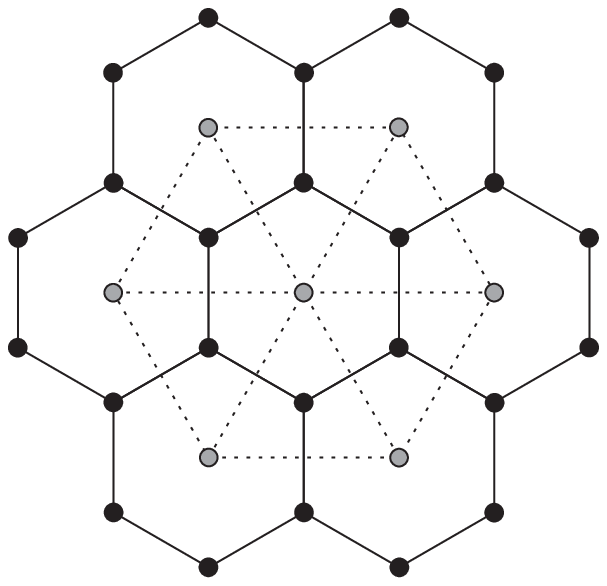}\\
{\small (a)}&{\small (b)}
\end{tabular}
\caption{Subgraphs of $B_{6\mathrm{e}}$ with subgraph of $D_{6\mathrm{e}}$ in 
the background (with dashed edges).}
\vspace{0.25in}
\label{d6e}
\end{figure}

Let us first see how Theorem~\ref{girth} simplifies the search for $B_{6\mathrm{e}}$. 
We begin by defining a graph $D_{6\mathrm{e}}$ such that $V(D_{6\mathrm{e}})
=\{C_1,\ldots,C_t\}$, where $C_1,\ldots,C_t$ are the $6$-CDC cycles of 
$B_{6\mathrm{e}}$, and $C_iC_j \in E(D_{6\mathrm{e}})$ if and only if $C_i$ and 
$C_j$ share an edge in $B_{6\mathrm{e}}$.\footnote{Although this definition of 
$D_{6\mathrm{e}}$ is very similar to that of a dual graph, we refrain from using 
this denomination because we do not assume that $B_{6\mathrm{e}}$ is planar.} 
Since $S_{6\mathrm{e}}$ is a subgraph of $B_{6\mathrm{e}}$, it induces the 
subgraph of $D_{6\mathrm{e}}$ shown in Figure~\ref{d6e}(a) with dashed edges. It 
is easy to see that $D_{6\mathrm{e}}$ is $6$-regular and that, by Lemma~\ref{3cycles},
every vertex in $B_{6\mathrm{e}}$ corresponds to a triangle in 
$D_{6\mathrm{e}}$ (and conversely).

Since $B_{6\mathrm{e}}$ is cubic, each triangle in $D_{6\mathrm{e}}$ shares an 
edge with exactly three other triangles and each edge in $D_{6\mathrm{e}}$ is 
shared by two triangles. These properties restrict the way in which the 
degree-$3$ vertices of $D_{6\mathrm{e}}$ in Figure~\ref{d6e}(a) may be connected 
to other vertices. In particular, they must not be connected among themselves, 
meaning that the six incomplete cycles of the subgraph of $B_{6\mathrm{e}}$ in the same figure 
must be completed as represented in Figure~\ref{d6e}(b), i.e., without further 
edge adjacency among themselves.

\begin{figure}[t]
\centering
\begin{tabular}{c@{\hspace{1.5cm}}c}
\includegraphics[scale=0.65]{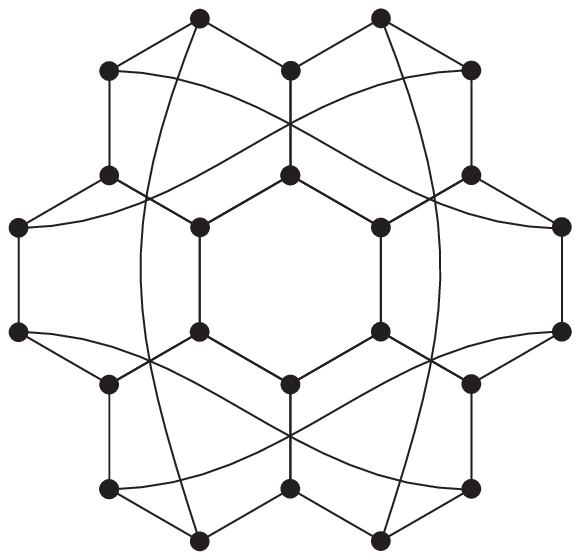}&
\includegraphics[scale=0.65]{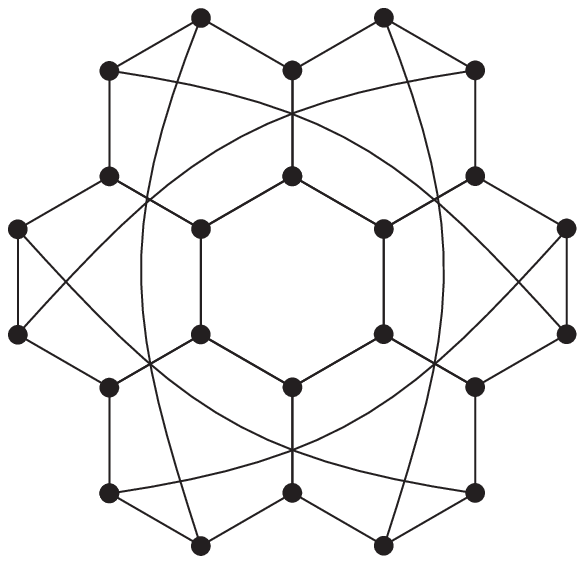}\\
{\small (a)}&{\small (b)}
\end{tabular}
\caption{Smallest girth-$6$ graphs, $B_{6\mathrm{e}}$ (a) and 
$B_{6\mathrm{e'}}$ (b), that have a $6$-CDC which contains every $6$-cycle.}
\vspace{0.25in}
\label{b6e}
\end{figure}

\begin{figure}[t]
\centering
\begin{tabular}{c@{\hspace{1.5cm}}c}
\includegraphics[scale=0.65]{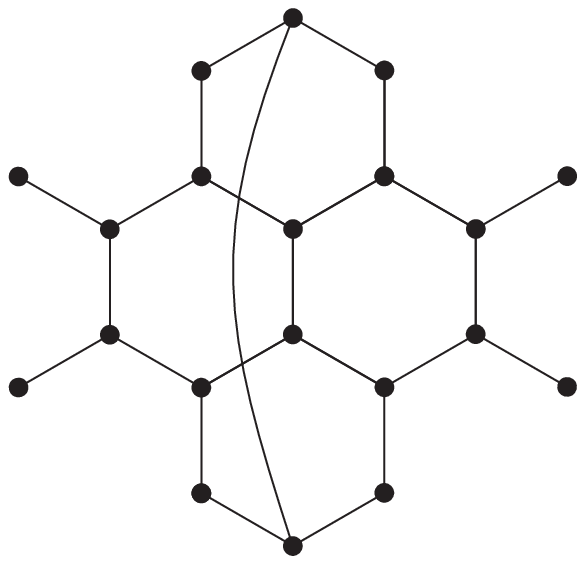}&
\includegraphics[scale=0.65]{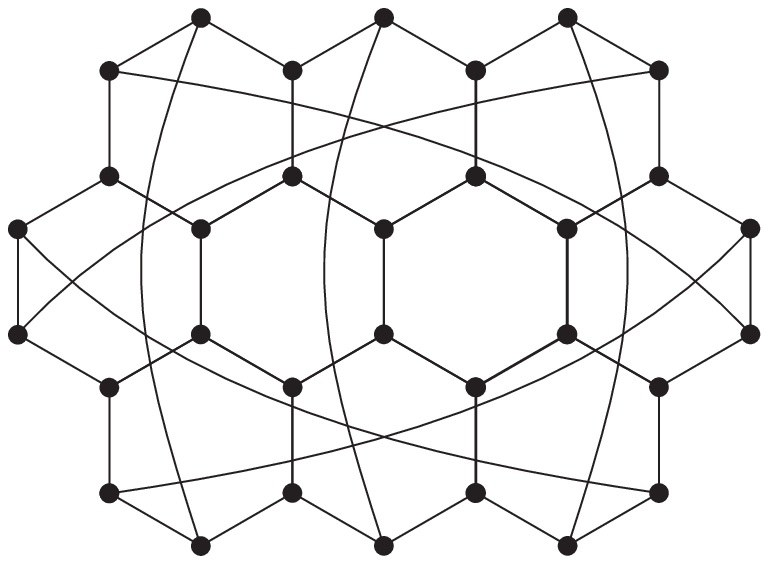}\\
{\small (a)}&{\small (b)}
\end{tabular}
\caption{$I_{6\mathrm{e}}$ (a) and the graph resulting from the first 
replacement of $S_{6\mathrm{e}}$ by $I_{6\mathrm{e}}$ in $B_{6\mathrm{e}}$ 
(b).}
\vspace{0.25in}
\label{i6e}
\end{figure}

It is relatively easy to see that the only
expansions of the graph of Figure~\ref{d6e}(b) that qualify as type-(iii)
instances of $B_6$ and moreover comply with Theorem~\ref{girth}
are the ones in Figure~\ref{b6e}. We denote them by 
$B_{6\mathrm{e}}$ (Figure~\ref{b6e}(a)) and $B_{6\mathrm{e'}}$ 
(Figure~\ref{b6e}(b)).\footnote{Unlike most of our previous illustrations, 
Figures~\ref{b6e} and \ref{i6e} contain no annotation for vertex or cycle 
identification. They are omitted for clarity and are furthermore needless, since 
in these figures all $6$-cycles are in the $6$-CDC.}  

The $I_{6\mathrm{e}}$ shown in Figure~\ref{i6e}(a) can be obtained 
straightforwardly from $S_{6\mathrm{e}}$ using the same restrictions as the ones 
used for obtaining $B_{6\mathrm{e}}$. Notice that the deficiency of 
$I_{6\mathrm{e}}$ is equal to that of $S_{6\mathrm{e}}$ ($12$ in both cases). 
Larger graphs in this class can be obtained in the same way as for the previous 
cases; the outcome of the first replacement of $S_{6\mathrm{e}}$ by 
$I_{6\mathrm{e}}$ in $B_{6\mathrm{e}}$ is shown in Figure~\ref{i6e}(b).  

\section{Completeness of the method and Hamiltonian cycles}
\label{complet}

It is possible to prove that the generation method discussed in
Sections~\ref{method} and \ref{applying} never outputs the same graph twice, and
also that all cubic graphs having a $6$-CDC are generated. In what follows, we
separate the $g<6$ case from that of $g=6$.

We first explain the absence of duplicates during generation.
Notice first that the method always keeps the girth constant as $I_g$ 
substitutes for $S_g$, so there is no interference between distinct-girth 
instances. In order to see that outputs are unique also for fixed $g$, consider
first the $g<6$ case. It then suffices to recall that all subgraphs isomorphic
to $S_g$ in $I_g$ or in $B_g-E(S_g)$ have the same cycle configuration, which
forbids any hybrid cycle configuration to be generated (i.e., a cycle
configuration with remnants from more than one $S_g$ instance).

For $g=6$, what might prevent the same simple argument from holding is that
there is, of course, the issue related to girth-$6$ graphs that we discussed in
Sections~\ref{method} and \ref{girth6}. In this case, the occurrence of more
than one cycle configuration for $S_g$-isomorphs is verified in all type-(i) and
(ii) instances of $B_g$. However, the extra cycle configurations always contain
a complete $6$-CDC cycle and our method never replaces them in the process of
generating new graphs from a type-(i) or (ii) $B_g$ instance. They only get
replaced when the method starts at a type-(iii) instance, so the no-duplicity
argument remains essentially unaltered.    

Let us now demonstrate that no cubic graph $G$ having a $6$-CDC is missed during
generation. The overall strategy here is to start from $G$ itself and to
repeatedly substitute $S_g$ for a subgraph of the current graph that is
isomorphic to $I_g$ until a $B_g$ instance is reached. If for every cubic $G$
that has a $6$-CDC we can argue that this ``reversal'' of the generation process
is possible, then we have proven that the method is complete.

Let us consider the $g<6$ case first. Let $G$ have girth $g$ and a $6$-CDC, and
recall that both $I_g$ and all instances of $B_g$ are girth-$g$ supergraphs of
$S_g$. The difference between them is that each $B_g$ is a cubic graph having a
$6$-CDC, while $I_g$ has nonzero deficiency and a cycle configuration with the
important property of being self-similar with respect to $S_g$. Because our
method relies on the explicit knowledge of every possible cycle configuration of
$S_g$, there are only two possibilities for $G$: either it is isomorphic to a
type-(i) instance of $B_g$ or it has a subgraph that is isomorphic to $I_g$. 
While in the former case $G$ is obviously generated by the method, in the latter
it is possible to recursively substitute $S_g$ for $I_g$ through a sequence of
ever smaller graphs until either a type-(i) instance of $B_g$ is finally 
obtained or else a substitution yields a graph that has less-than-$g$ girth.
If it is not the case that the process ends at a type-(i) instance of $B_g$,
then by definition the last substitution must have been applied on a type-(ii)
instance of $B_g$. We then conclude that, in any case, $G$ is output by the
method. 

The case of $g=6$ is analogous, but the possibilities for ending the sequence of
substitutions are more varied. The sequence may end when a type-(i) instance of
$B_g$ is reached, or when a graph is obtained whose girth is less than $g$ (if
$G$ has $6$-cycles that are not in the $6$-CDC), or yet when a graph is obtained
that has acquired $6$-cycles that are not in the $6$-CDC (if all of $G$'s
$6$-cycles are in the $6$-CDC). Similarly to the case of $g<6$, if the process
does not end at a type-(i) instance of $B_g$, then by definition the last
substitution must have been applied respectively on a type-(ii) or (iii)
instance of $B_g$. Once again, $G$ is in any case seen to be output by the
method. 

It is important to note that this argument for the method's completeness relies
crucially on the fact that every possible instance of $B_g$ is known. For
$g=3,\ldots,6$, this is part of what we established in Sections~\ref{girth3}
through \ref{girth6}. A key observation related to our exhaustive enumeration of
$B_g$ instances in those sections is that in none of those instances is
more than one cycle configuration of $S_g$ present, with the important exception
in the girth-$6$ case we noted in Section~\ref{method}. For this reason, in the
above completeness argument we need not concern ourselves with the presence
of multiple cycle configurations for $S_g$: in girth-$g$ cubic graphs that
have a $6$-CDC, such multiplicity never occurs, unless $g=6$ and the cycle
configuration of $S_g$ does not include a complete $6$-CDC cycle---in this case,
the argument is already split into finishing at a type-(ii) or a type-(iii)
instance of $B_g$.

\begin{figure}[t]
\centering
\begin{tabular}{c@{\hspace{0.5cm}}c@{\hspace{0.5cm}}c@{\hspace{1.0cm}}c@{\hspace{1.5cm}}c}
\includegraphics[scale=0.65]{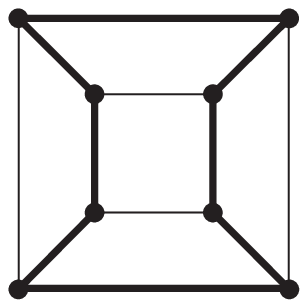}&
\includegraphics[scale=0.65]{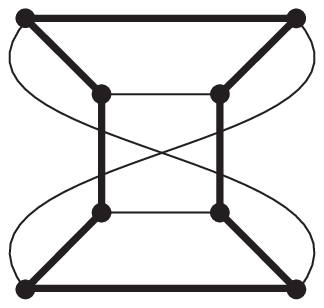}&
\includegraphics[scale=0.65]{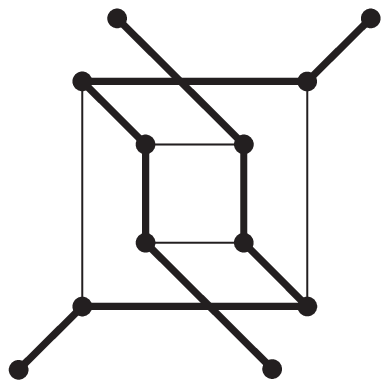}&
\includegraphics[scale=0.65]{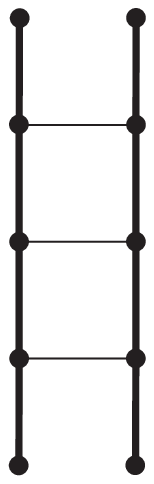}&
\includegraphics[scale=0.65]{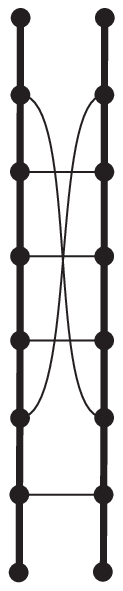}\\
{\small (a)}&{\small (b)}&{\small (c)}&{\small (d)}&{\small (e)}
\end{tabular}
\caption{Hamiltonian cycle in $B_{4\mathrm{a}}$, $B_{4\mathrm{b}}$, or
$B_{4\mathrm{c}}$ (a), and in $B_{4\mathrm{b'}}$ (b), and equivalent paths in 
$I_{4\mathrm{a}}$ (c), $I_{4\mathrm{b}}$ (d), and $I_{4\mathrm{c}}$ (e).}
\vspace{0.25in}
\label{hamilt4}
\end{figure}

\begin{figure}[p]
\centering
\begin{tabular}{c@{\hspace{1.5cm}}c}
\includegraphics[scale=0.65]{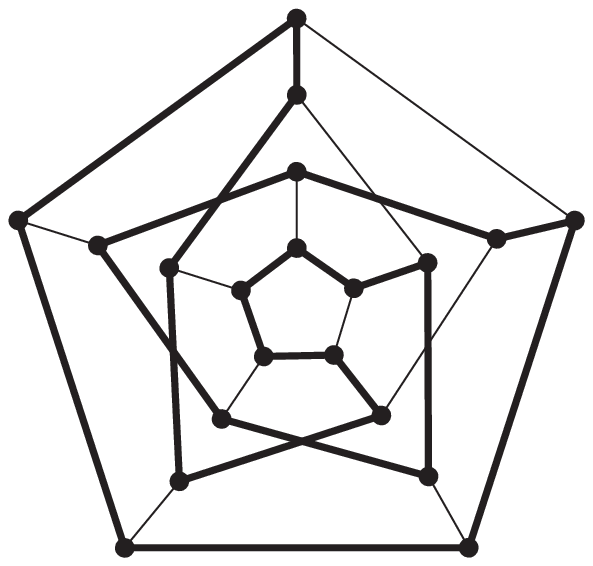}&
\includegraphics[scale=0.65]{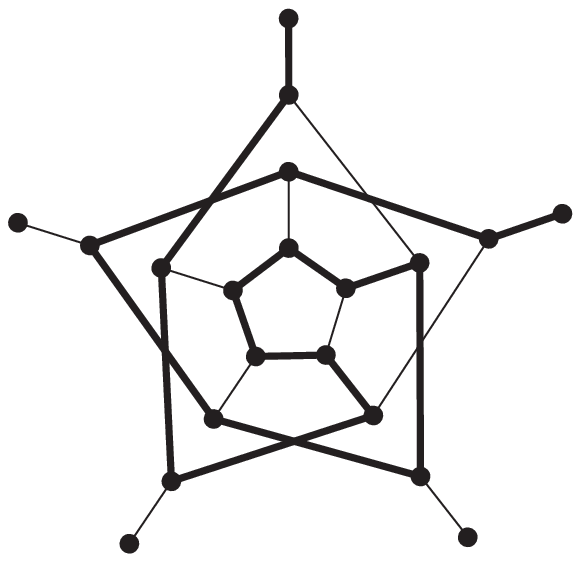}\\
{\small (a)}&{\small (b)}
\end{tabular}
\caption{Hamiltonian cycle in $B_{5\mathrm{c}}$ (a) and equivalent path in 
$I_{5\mathrm{c}}$ (b).}
\vspace{0.25in}
\label{hamilt5}
\end{figure}

\begin{figure}[p]
\centering
\begin{tabular}{c@{\hspace{1.5cm}}c}
\includegraphics[scale=0.65]{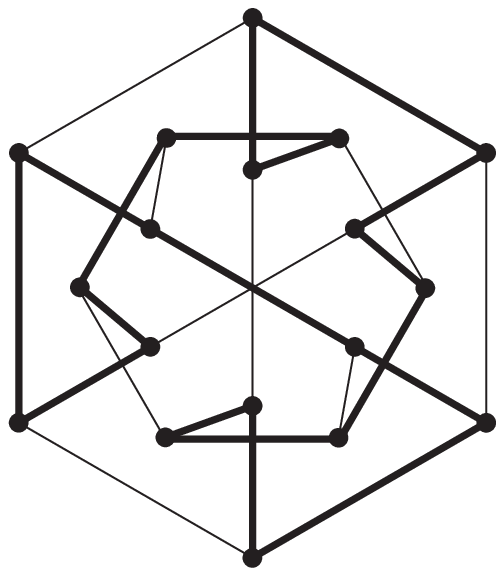}&
\includegraphics[scale=0.65]{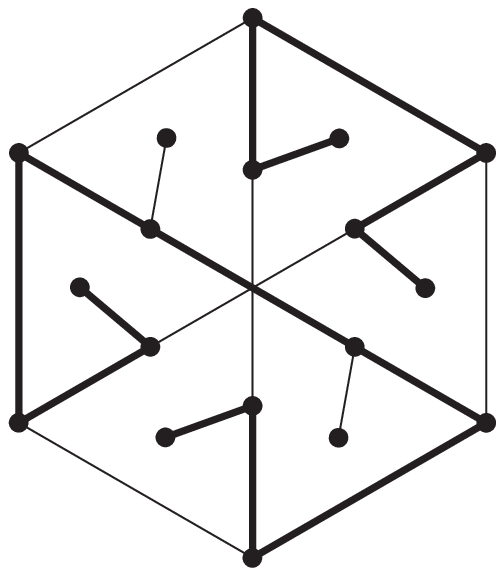}\\
{\small (a)}&{\small (b)}
\end{tabular}
\caption{Hamiltonian cycle in $B_{6\mathrm{a}}$ (a) and equivalent paths in 
$I_{6\mathrm{a}}$ (b).}
\vspace{0.25in}
\label{hamilt6a}
\end{figure}

\begin{figure}[p]
\centering
\begin{tabular}{c@{\hspace{1.0cm}}c@{\hspace{1.0cm}}c}
\includegraphics[scale=0.65]{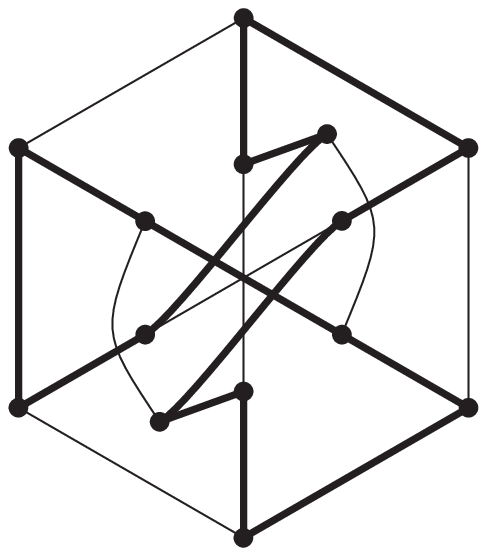}&
\includegraphics[scale=0.65]{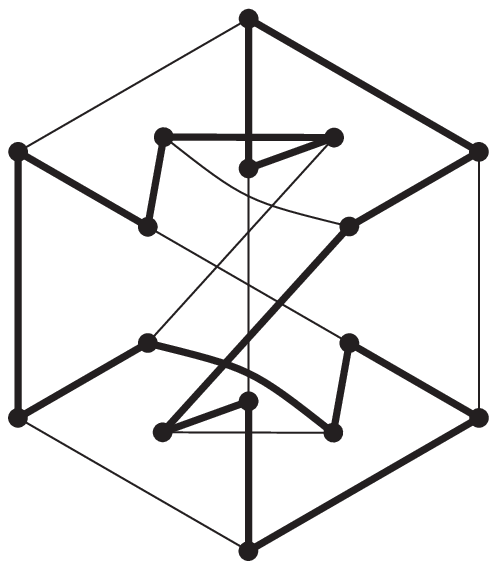}&
\includegraphics[scale=0.65]{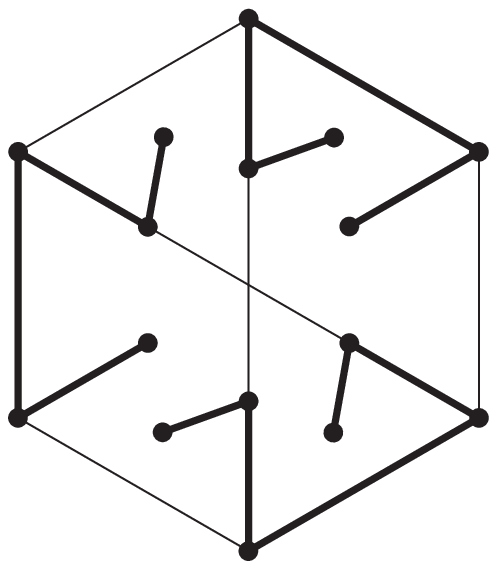}\\
{\small (a)}&{\small (b)}&{\small (c)}
\end{tabular}
\caption{Hamiltonian cycle in $B_{6\mathrm{b}}$ (a) and in 
$B_{6\mathrm{b'}}$ (b), and equivalent paths in $I_{6\mathrm{b}}$ (c).}
\vspace{0.25in}
\label{hamilt6b}
\end{figure}

\begin{figure}[p]
\centering
\begin{tabular}{c@{\hspace{1.5cm}}c}
\includegraphics[scale=0.65]{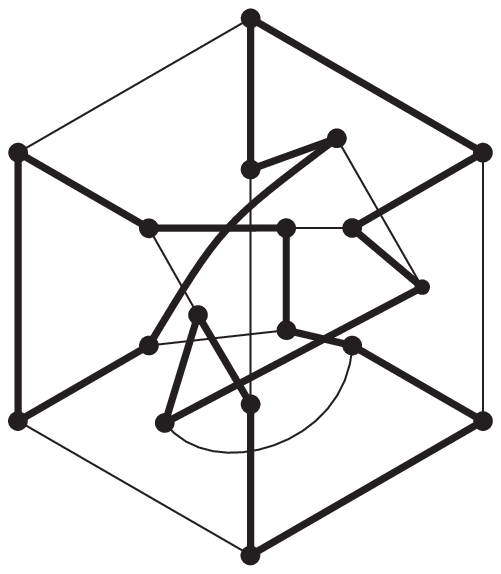}&
\includegraphics[scale=0.65]{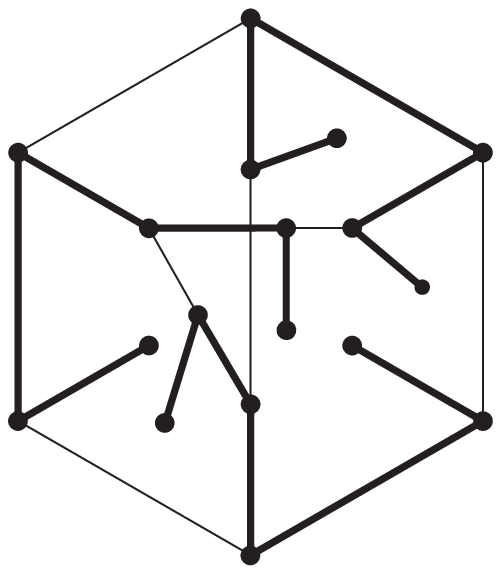}\\
{\small (a)}&{\small (b)}
\end{tabular}
\caption{Hamiltonian cycle in $B_{6\mathrm{c}}$ (a) and equivalent paths in 
$I_{6\mathrm{c}}$ (b).}
\vspace{0.25in}
\label{hamilt6c}
\end{figure}

\begin{figure}[p]
\centering
\begin{tabular}{c@{\hspace{0.5cm}}c@{\hspace{0.5cm}}c}
\includegraphics[scale=0.65]{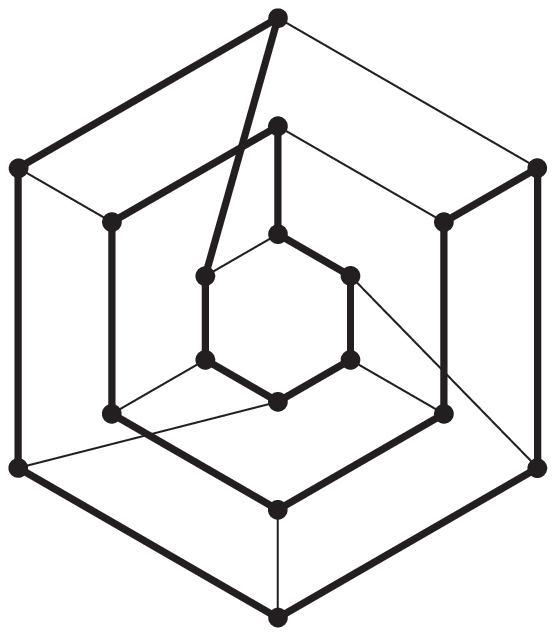}&
\includegraphics[scale=0.65]{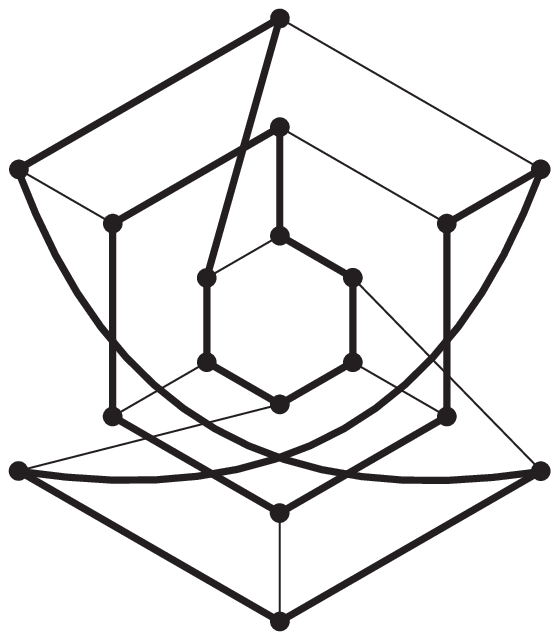}&
\includegraphics[scale=0.65]{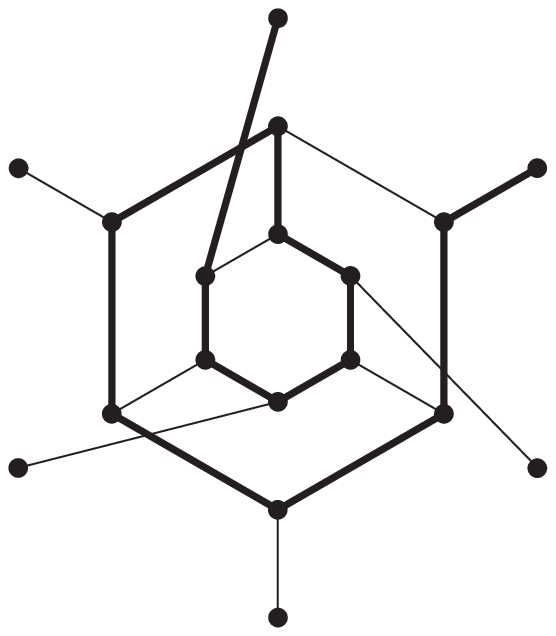}\\
{\small (a)}&{\small (b)}&{\small (c)}
\end{tabular}
\caption{Hamiltonian cycle in $B_{6\mathrm{d}}$ (a) and in $B_{6\mathrm{d'}}$ 
(b), and equivalent path in $I_{6\mathrm{d}}$ (c).}
\vspace{0.25in}
\label{hamilt6d}
\end{figure}

\begin{figure}[p]
\centering
\begin{tabular}{c@{\hspace{1.5cm}}c}
\includegraphics[scale=0.65]{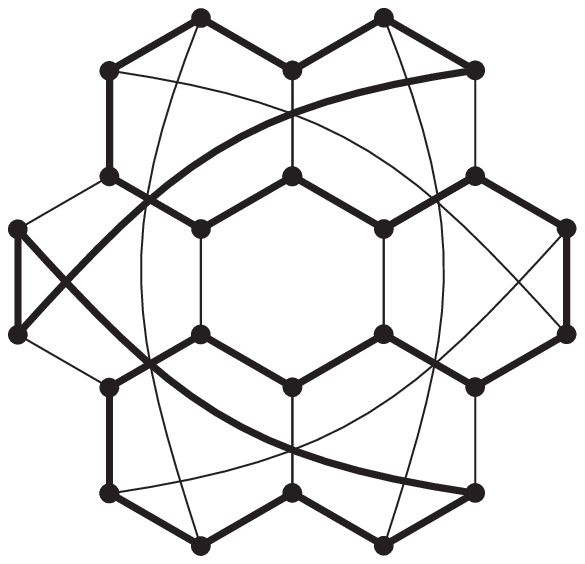}&
\includegraphics[scale=0.65]{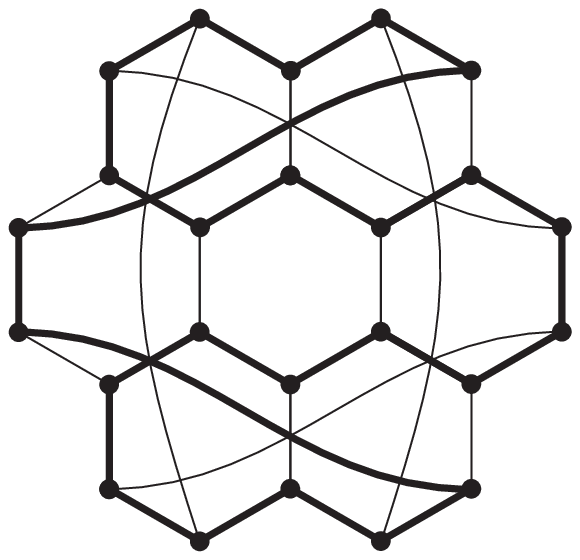}\\
{\small (a)}&{\small (b)}
\vspace{0.5cm}
\end{tabular}
\begin{tabular}{c}
\includegraphics[scale=0.65]{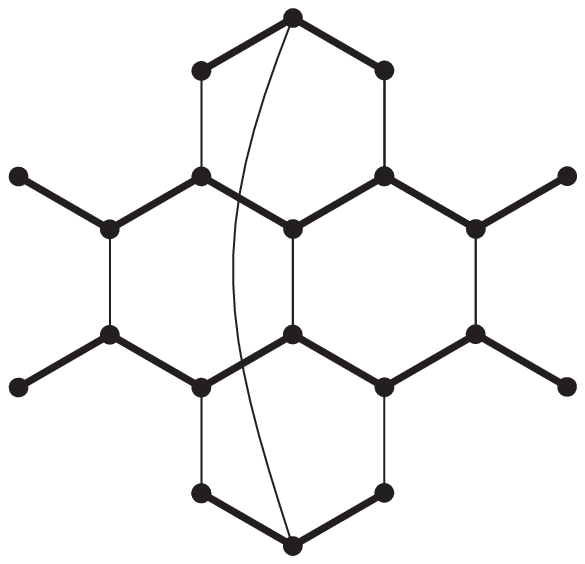}\\
{\small (c)}
\end{tabular}
\caption{Hamiltonian cycle in $B_{6\mathrm{e}}$ (a) and in $B_{6\mathrm{e'}}$ 
(b), and equivalent paths in $I_{6\mathrm{e}}$ (c).}
\vspace{0.25in}
\label{hamilt6e}
\end{figure}

It is also possible to identify a Hamiltonian cycle in every cubic graph that 
has a $6$-CDC. We first find a Hamiltonian cycle $C$ in $B_g$. Then we take 
paths in $I_g$ that are equivalent to the path used by $C$ in the $S_g$-isomorphic subgraph 
of $B_g$. For example, in $B_3$ we have the Hamiltonian cycle $dacbefd$, whose 
intersection with $S_3$ is the path $dacbe$ (cf.\ Figure~\ref{b3&s3}), and an 
equivalent path in $I_3$ (cf.\ Figure~\ref{I3&1iter}(a)) is $dglkjabchie$. 
Successive substitutions of $I_3$ for $S_3$ will then always ensure the presence 
of a Hamiltonian cycle. In Figures~\ref{hamilt4} through \ref{hamilt6e} we show 
(as thick edges) Hamiltonian cycles in $B_g$ and equivalent paths in $I_g$ for 
all the remaining pertinent values of $g$.

\section{$\mathbf{6}$-CDC's and the minimal chordal sense of direction}
\label{mcsd}

In this section, $G$ is no longer assumed to have a $6$-CDC, but rather to be 
such that every one of its edges has two labels, each corresponding to one of 
its end vertices. In \cite{santoro}, a property of this edge labeling has been 
introduced which can considerably reduce the complexity of many problems in 
distributed computing \cite{flocchini7}. This property refers to the ability of 
a vertex to distinguish among its incident edges according to a globally 
consistent scheme and is formally described in \cite{flocchini1}. An edge 
labeling for which the property holds is called a \emph{sense of direction}. 
It is \emph{symmetric} if for every edge one label can be inferred from
the other. While the full-fledged definition of sense of direction is irrelevant
to our present discussion, the special sense of direction that we describe next
is closely related to a cubic graph's having a $6$-CDC.

\begin{figure}[t]
\centering
\includegraphics[scale=0.65]{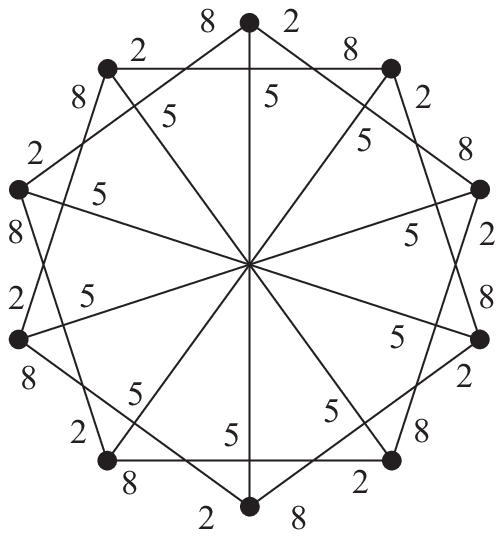}
\caption{A graph with an edge labeling that is an MCSD. Vertices are ordered 
clockwise.}
\vspace{0.25in}
\label{example}
\end{figure}

We say that a sense of direction is \emph{minimal} if it requires exactly 
$\Delta(G)$ distinct labels, where $\Delta(G)$ is the maximum degree in $G$.
A particular instance of symmetric sense of direction, called a \emph{chordal
sense of direction}, can be constructed on any graph by fixing an arbitrary 
cyclic ordering of the vertices and, for each edge $uv$, selecting the 
difference (modulo $n$) from the rank of $u$ in the ordering to that of $v$ as
the label of $uv$ that corresponds to $u$ (likewise, the label that corresponds
to $v$ is the rank difference from $v$ to $u$). In Figure~\ref{example}, an 
example is given of a minimal chordal sense of direction (MCSD). For a survey on
sense of direction, we refer the reader to \cite{flocchini8}.

Before proceeding to our result in this section, we pause briefly to review 
some relevant definitions. Given a finite group $A$ and a set of generators 
$S\subseteq A$, a \emph{Cayley graph} is a graph $H$ whose vertices are the 
elements of the group ($V(H)=A$) and whose edges correspond to the action of the 
generators ($uv \in E(H) \iff \exists s \in S: v=s*u$, where $*$ is the 
operation defined for $A$). We assume that the set of generators is closed under 
inversion, so $H$ is an undirected graph. A \emph{circulant graph} (also known 
as a \emph{chordal ring}) is a Cayley graph over $\mathbb{Z}_n$, the cyclic 
group of order $n$ under the addition operation.

In \cite{mcsd}, we have analyzed the class of regular graphs that 
admit an MCSD, and proved that this class is equivalent to that of circulant 
graphs. In this section, we prove that nearly all the cubic graphs in this class 
have a $6$-CDC, the only exception being $K_4$. We start by noting that a 
characterization of such cubic graphs follows directly from the results of 
\cite{mcsd} (specifically, Theorem~5, Lemma~6, and Lemma~7) and can be 
stated as follows.

\begin{lem}
\label{equivalence}
$G$ admits an MCSD (or, equivalently, $G$ is a circulant graph) if and only if 
$G$ is isomorphic to either $M_n$, with $n\geq 4$, or to $T_{n,2}$, with 
$n\geq 6$.
\end{lem}

We now present the main result of this section.

\begin{thm}
\label{mcsd&6cdc}
Except for $K_4$, every cubic graph that admits an MCSD has a $6$-CDC. 
\end{thm}
\begin{proof}
By Lemma~\ref{equivalence}, it suffices to consider instances of $M_n$, with 
$n\geq 4$, and of $T_{n,2}$, with $n\geq 6$.

First, notice that all instances of $M_n$ or $T_{n,2}$ have girth less than $5$. 
Also, the only instances of girth $3$ are $M_4$ and $T_{6,2}$. The former of 
these is isomorphic to $K_4$ and obviously does not have a $6$-CDC. As for the 
latter, a $6$-CDC is shown in Figure~\ref{b3&s3}(a).

Let us then consider the girth-$4$ instances; we do this by resorting to the 
material of Section~\ref{girth4}. First notice that $B_{4\mathrm{b'}}$ is 
isomorphic to $M_6$ and that all instances of $M_n$, for $n>6$, can be generated 
by successive replacements of $S_{4\mathrm{b'}}$ by $I_{4\mathrm{b'}}$ in 
$B_{4\mathrm{b'}}$. Similarly, $B_{4\mathrm{b}}$ is isomorphic to $T_{8,2}$ and 
all instances of $T_{n,2}$, for $n>8$, can be generated by successive 
replacements of $S_{4\mathrm{b}}$ by $I_{4\mathrm{b}}$ in $B_{4\mathrm{b}}$.  
It then follows that every instance of $M_n$ or $T_{n,2}$
having $n\geq 6$ also has a $6$-CDC. 
\end{proof} 

\section{Conclusions}
\label{conclusions}

We have in this paper demonstrated how to generate all the cubic graphs that 
have a $6$-CDC in a constructive manner. For an arbitrary cubic graph $G$ with 
$n$ vertices and girth $g$, our method provides, at least in principle, a 
mechanism for checking whether $G$ has a $6$-CDC: one simply generates all 
girth-$g$ cubic graphs on $n$ vertices that have a $6$-CDC and checks each one 
against $G$ for isomorphism. This check, we recall, can be performed 
polynomially for cubic graphs \cite{luks}.

Our method also provides a mechanism for pinpointing a Hamiltonian cycle in any 
cubic graph that has a $6$-CDC. That all such graphs are Hamiltonian is a result 
consistent with the one in \cite{mcsd}, given our further demonstration, in this 
paper, that all non-$K_4$ cubic graphs that have an MCSD also have a $6$-CDC. 
The alluded result in \cite{mcsd} is that all regular graphs that have an MCSD 
are Hamiltonian.

Our results relating $6$-CDC's to MCSD's in cubic graphs create a bridge 
connecting these two concepts and also, by consequence, the notion of a 
circulant graph in the cubic case. The obvious implication of this is that 
results obtained within one context can now be extended directly to any other.

There are several open problems that may be addressed to expand on the results
we have presented. Some of them come from generalizing the vertices' fixed
degree or the constant length of a CDC's cycles, or yet from relaxing at least
one of the two constraints by letting vertices have different degrees or CDC
cycles different lengths. Relaxing both is really tantamount to addressing the
Szekeres-Seymour conjecture \cite{szekeres,seymour}, according to which every
$2$-edge-connected graph has a CDC. This conjecture has stood for over thirty
years, so perhaps an easier (though by no means trivial) starting problem for
further research is to characterize the $k$-regular graphs that have a $2k$-CDC,
$k\ge 3$.

\subsection*{Acknowledgments}
                                                                                
The authors acknowledge partial support from CNPq, CAPES, and a FAPERJ BBP 
grant. They also thank C. Thomassen for pointing references 
\cite{altshuler,thomassen2,thomassen1} to them.

\bibliographystyle{plain}
\bibliography{rleao}

\begin{thebibliography}{10}

\bibitem{altshuler}
A.~Altshuler.
\newblock Construction and enumeration of regular maps on the torus.
\newblock {\em Discrete Mathematics}, 4:201--217, 1973.

\bibitem{bermond}
J.~C. Bermond, F.~Cornellas, and D.~F. Hsu.
\newblock Distributed loop computer networks: a survey.
\newblock {\em Journal of Parallel and Distributed Computing}, 24:2--10, 1995.

\bibitem{bondy}
J.~A. Bondy and U.~S.~R. Murty.
\newblock {\em Graph Theory with Applications}.
\newblock North-Holland, New York, NY, 1976.

\bibitem{deza}
M.~Deza, P.~W. Fowler, A.~Rassat, and K.~M. Rogers.
\newblock Fullerenes as tilings of surfaces.
\newblock {\em Journal of Chemical Information and Computer Sciences},
  40:550--558, 2000.

\bibitem{flocchini7}
P.~Flocchini, B.~Mans, and N.~Santoro.
\newblock On the impact of sense of direction on message complexity.
\newblock {\em Information Processing Letters}, 63:23--31, 1997.

\bibitem{flocchini1}
P.~Flocchini, B.~Mans, and N.~Santoro.
\newblock Sense of direction: definitions, properties and classes.
\newblock {\em Networks}, 32:165--180, 1998.

\bibitem{flocchini8}
P.~Flocchini, B.~Mans, and N.~Santoro.
\newblock Sense of direction in distributed computing.
\newblock {\em Theoretical Computer Science}, 291:29--53, 2003.

\bibitem{kirby}
E.~C. Kirby, R.~B. Mallion, and P.~Pollack.
\newblock Toroidal polyhexes.
\newblock {\em Journal of the Chemical Society Faraday Transactions},
  89:1945--1953, 1993.

\bibitem{mcsd}
R.~S.~C. Le{\~a}o and V.~C. Barbosa.
\newblock Minimal chordal sense of direction and circulant graphs.
\newblock In R.~Kr\'{a}lovi\v{c} and P.~Urzyczyn, editors, {\em Mathematical
  Foundations of Computer Science 2006}, volume 4162 of {\em Lecture Notes in
  Computer Science}, pages 670--680, Berlin, Germany, 2006. Springer-Verlag.

\bibitem{luks}
E.~M. Luks.
\newblock Isomorphism of graphs of bounded valence can be tested in polynomial
  time.
\newblock {\em Journal of Computer and System Sciences}, 25:42--65, 1982.

\bibitem{santoro}
N.~Santoro.
\newblock Sense of direction, topological awareness and communication
  complexity.
\newblock {\em SIGACT News}, 2:50--56, 1984.

\bibitem{seymour}
P.~D. Seymour.
\newblock Sums of circuits.
\newblock In J.~A. Bondy and U.~S.~R. Murty, editors, {\em Graph Theory and
  Related Topics}, pages 341--355. Academic Press, New York, NY, 1979.

\bibitem{szekeres}
G.~Szekeres.
\newblock Polyhedral decompositions of cubic graphs.
\newblock {\em Bulletin of the Australian Mathematical Society}, 8:367--387,
  1973.

\bibitem{thomassen2}
C.~Thomassen.
\newblock Tilings of the torus and the {K}lein bottle and vertex-transitive
  graphs on a fixed surface.
\newblock {\em Transactions of the American Mathematical Society},
  323:605--635, 1991.

\bibitem{thomassen1}
C.~Thomassen.
\newblock Triangulating a surface with a prescribed graph.
\newblock {\em Journal of Combinatorial Theory, Series B}, 57:196--206, 1993.

\end{thebibliography}

\end{document}